\documentclass[twocolumn,twocolappendix]{aastex631}
\usepackage[T1]{fontenc}
\usepackage{amsmath,mathtools,mathrsfs,amssymb}
\usepackage{hyperref}
\usepackage{verbatim}
\usepackage{xspace}
\usepackage{graphicx}
\usepackage{longtable}
\usepackage{upgreek}

\newcommand{\bhspin}{a_*}
\newcommand{\rhigh}{r_{\mathrm{high}}}
\newcommand{\rlow}{r_{\mathrm{low}}}
\newcommand{\sgra}{{Sgr\,A*}\xspace}
\newcommand{\kB}{k_{\rm B}}

\newcommand{\sigmac}{\sigma_{\rm c}}

\newcommand{\Ti}{T_{\rm i}}
\newcommand{\Te}{T_{\rm e}}
\newcommand{\ui}{u_{\rm i}}
\newcommand{\ue}{u_{\rm e}}

\newcommand{\iharm}[0]{{\tt{iharm3d}}}
\newcommand{\ipole}[0]{{\tt{ipole}}\xspace}

\renewcommand{\apjl}{Astrophys.~J.~Lett.}

\renewcommand{\apjs}{Astrophys.~J.~Supp.}

\renewcommand{\aap}{Astron.~Astro.}

\begin{document}

\shorttitle{The Helicity Barrier in Black Hole Accretion}
\shortauthors{Wong \& Arzamasskiy}

\correspondingauthor{George~N.~Wong}
\email{gnwong@ias.edu}

\title{Balanced Turbulence and the Helicity Barrier in Black Hole Accretion}

\author[0000-0001-6952-2147]{George~N.~Wong}
\affiliation{School of Natural Sciences, Institute for Advanced Study, 1 Einstein Drive, Princeton, NJ 08540, USA}
\affiliation{Princeton Gravity Initiative, Princeton University, Princeton, New Jersey 08544, USA}

\author[0000-0002-5263-9274]{Lev~Arzamasskiy}
\affiliation{School of Natural Sciences, Institute for Advanced Study, 1 Einstein Drive, Princeton, NJ 08540, USA}

\begin{abstract}
Horizon-scale observations from the Event Horizon Telescope (EHT) have enabled precision study of supermassive black hole accretion. Contemporary accretion modeling often treats the inflowing plasma as a single, thermal fluid, but microphysical kinetic effects can lead to significant deviations from this idealized picture. We investigate how the helicity barrier influences EHT-accessible electromagnetic observables by employing a simple model for electron heating based on kinetic physics and the cascade of energy and helicity in unbalanced turbulence. Although the helicity barrier plays only a minor role in regions with high plasma-$\beta$, like in SANE disks, it may substantially impact in regions with more ordered magnetic fields, such as the jet and its surrounding wind in SANE flows as well as throughout the entire domain in MAD flows. In SANE flows, emission shifts from the funnel wall towards the lower-magnetization disk region; in MAD flows the emission morphology remains largely unchanged. Including the helicity barrier leads to characteristically lower electron temperatures, and neglecting it can lead to underestimated accretion rates and inferred jet powers. The corresponding higher plasma densities result in increased depolarization and Faraday depths thereby decreasing the amplitude of the $\beta_2$ coefficient while leaving its angle unchanged. Both the increased jet power and lower $\left|\beta_2\right|$ may help alleviate outstanding tensions between modeling and EHT observations. We also find that the estimated ring diameter may be underestimated when the helicity barrier is neglected. Our results underscore the significance of the helicity barrier in shaping black hole observables and inferred accretion system parameters.
\end{abstract}

\keywords{plasma astrophysics (1261) -- magnetohydrodynamics (1964) -- radiative transfer (1335) -- accretion (14)}

\section{Introduction}

Low luminosity active galactic nuclei (LLAGN) are usually modeled as radiatively inefficient accretion flows (RIAFs) onto supermassive black holes \citep[e.g.,][]{ichimaru_1977_BimodalBehaviorAccretion,rees_1982_IonsupportedToriOrigin,narayan_1995_ADAF,reynolds_1996_QuiescentBlackHole}. In contrast to radiatively efficient thin disks model, RIAFs comprise geometrically thick, optically thin disks of hot plasma that circle the hole at subkeplerian speeds; in RIAFs, the gravitational binding energy of the inflowing plasma is converted into heat that cannot be radiated away before the plasma accretes down to the horizon. The excess heat provides a thermal pressure that supports a puffy disk. 

Two of the LLAGN with the largest known sizes on the sky lie at the centers of our galaxy (\sgra) and the nearby elliptical galaxy Messier 87 (M87*). These two presumed RIAF sources are large enough to be directly observed by the Event Horizon Telescope (EHT), and the EHT's very long baseline interferometric experiment has produced radio images of the horizon-scale emission in spectacular detail \citep{doeleman_2012_JetLaunchingStructureResolved,akiyama_2015_230GHzVLBI,akiyama_2017_ImagingSchwarzschildradiusscaleStructure,EHTC_2019_1,EHTC_2021_7,EHTC_2022_1}. The observations can be used to probe plasma physics in these extreme environments while also providing constraints on key physical parameters of the accretion system, like the black hole mass and spin, the system accretion rate, and the amount of magnetic flux trapped on the horizon \citep{EHTC_2019_5,EHTC_2021_8,EHTC_2022_5}.  
Ongoing measurements and the next generation of these experiments will help inform further parameter constraints and enable precision tests of our understanding of the physics governing these systems. The first observational results have already revealed tension between models and data in quantities like the resolved linear polarization fraction, which magnetized models often overproduce, and the predicted jet powers, which lie almost categorically at the lower end of the observational bounds.

Parameter estimation infers that M87* and \sgra are likely Coulomb collisionless since the path length to a Coulomb interaction greatly exceeds the size of the system. The electrons and ions that make up the infalling plasma therefore do not have time to equilibrate and relax to a thermal Maxwell--J\"{u}ttner distribution \citep[e.g.,][]{shapiro_1976_twotemp,mahadevan_1997_AreParticlesAdvectiondominated,quataert_1998_ParticleHeatingAlfvenic,sadowski_2017_RadiativeTwotemperatureSimulations,chael_2018_RoleElectronHeating,ryan_2018_TwotemperatureGRRMHDSimulations}. Nevertheless, it may be that the ion- and electron-distribution functions are independently thermal, since intraspecies interactions due to kinetic plasma instabilities can drive particle-wave interactions that enable relaxation (see \citealt{kunz_2014_FirehoseMirrorInstabilities} and discussion therein). 

The mechanisms that govern the heating and cooling of the ions and electrons are the subject of detailed study. Since the radio emission observed from \sgra and M87* is produced by the synchrotron process, the distribution function of the electrons plays a crucial role in determining the observational features of the sources.
Accurately modeling particle acceleration is thus essential, since turbulent heating, reconnection, and shocks all yield different heating profiles and it is likely that different combinations of all heating mechanisms operate in different parts of the accretion flow. Interpreting the nonthermal features of the observations will also require a detailed understanding of the processes that determine the local particle distribution functions. 

\vspace{0.5em}

RIAFs have long been modeled with semianalytic \citep[e.g.,][]{ichimaru_1977_BimodalBehaviorAccretion,rees_1982_IonsupportedToriOrigin,narayan_1995_ADAF,ozel_2000_HybridThermalNonthermalSynchrotron} and numerical \citep[e.g.,][]{hawley_2000_GlobalMagnetohydrodynamicalSimulations,devilliers_2003_grmhdtori,mckinney_2004_harmbz} methods.
The latter models are usually produced through general relativistic magnetohydrodynamics (GRMHD) simulation and are often favored over semianalytic models because they naturally incorporate properties of the turbulent dynamics, produce variability, and effectuate the connection between the accretion disk, wind, and jet, all of which may play an important role in determining the observational appearance of the system. The output of the fluid simulations is typically processed through general relativistic ray tracing (GRRT) codes to generate simulated observables like images and spectra. 

In the standard modeling procedure, only the total energy of the fluid is tracked and evolved in the simulation. Since the electron distribution function is required to compute the radiative transfer coefficients, the typical modeling approach is to assume that the electrons are thermal and assign their temperature in a post-processing step by partitioning the total internal energy of the fluid into the ions and electrons following a prescription that depends on the ratio of the gas-to-magnetic pressure and the magnetization. These thermodynamic prescriptions are usually motivated by (kinetic) plasma theory for turbulent cascades, magnetic reconnection, collisionless shocks, and so on.\footnote{Some alternative two-temperature methods track the internal energies of the ions and electrons separately and model electron heating as some fraction of the total numerical dissipation \citep[e.g.,][]{ressler_2015_ElectronThermodynamicsGRMHD,sadowski_2017_RadiativeTwotemperatureSimulations}.} This approach introduces significant uncertainty and may well explain the model/data tension: the presence of a population of cold electrons would require an increased mass accretion rate, higher jet power, and would result in more depolarization from Faraday scrambling.

\vspace{0.5em}

The turbulent cascade model is often invoked to quantify the energy partition: energy is injected at large scales (e.g., from the magnetorotational instability or large-scale torques) and cascades to higher wavenumbers until it is dissipated as thermal energy into the ions or electrons at their associated Larmor scales. But when the turbulence is imbalanced, not all of the energy injected at large scales can be treated the same way, and if plasma $\beta\equiv P_{\rm gas}/P_{\rm mag}$ is small, conservation of helicity can inhibit energy flow in the cascade.

In low-$\beta$ plasmas, the sense of the helicity cascade above and below the ion Larmor scale changes: as wavenumber increases into the kinetic scale, the fluid cross-helicity transforms conservatively into magnetic helicity and the direction of the helicity cascade inverts \citep{biskamp_2003_mhdturbulence,Meyrand2021,Squire2022}. Since (generalized) helicity is conserved, the helicity-endowed component of the turbulence is then unable to cascade below the ion Larmor scale and an effective helicity barrier is produced, limiting the fraction of the energy at large scales that can reach and heat the electrons. Observational evidence for the helicity barrier's operation has recently been found in the context of the solar wind via correlation of ion-cyclotron waves and electron-scale turbulence \citep{bowen_2023_MediationCollisionlessTurbulentDissipation}.

In this paper, we use a simple model to study the effect of the helicity barrier on electron heating in radiatively inefficient accretion flows and probe its effect on the electromagnetic observables accessible to the horizon-scale radio observations. Since the presence of long-lived imbalanced turbulence is required for the helicity barrier to operate, one might expect the helicity barrier to be most important in the directed winds above the surface of the disk, but it is challenging to generate a predictive model for the quantitative details. In this study, we consider a limited subset of models and perform a preliminary study of the effect of the helicity barrier. We thus aim only to identify and describe broad qualitative trends to judge the importance of the effect and whether it may help explain contemporary questions raised by the data. We leave a more detailed study to future work.

\vspace{0.5em}

In Section~\ref{sec:heating} we provide a brief overview of the problem of particle heating and its connection to the helicity barrier. We describe the details of our numerical methods in Section~\ref{sec:numerical}. In Section~\ref{sec:results}, we present the results of our numerical exploration, and we discuss model assumptions and limitations in Section~\ref{sec:discussion}. We conclude in Section~\ref{sec:summary}.

\section{Particle Heating in Collisionless Turbulence}
\label{sec:heating}

Robust interpretation of black hole images requires a detailed understanding of how emission is produced by the accretion flow. Most of the emission is due to synchrotron radiation \citep[e.g.,][]{yuan_2014_HotAccretionFlows} and thus it is very sensitive to the electron momentum distribution function. The electron distribution function is determined by the details of the particle heating mechanisms that transform the gravitational energy released during accretion into thermal kinetic energy. The channels responsible for this conversion include dissipation in shocks \citep[e.g.,][]{tidman_1971_collisionlessshocks,blandford_1987_shockacceleration,ghavamian_2007_tpte_clshocks,mondal_2020_shocksaccretion,tran_2020_eheating_lowbetashocks}, at reconnection sites \citep[e.g.,][]{bisnovatyikogan_1997_OhmicADAF,rowan_2017_ElectronProtonHeating,werner_2018_NonthermalParticleAcceleration,rowan_2019_ElectronProtonHeating,sironi_2020_KineticReconnection,werner_2021_3dReconnectionPairs}, and in turbulent cascades. The relative importance of these channels is not well understood. In this paper, we focus on the turbulent cascade as the main source of energy for electrons. 

\subsection{The turbulent cascade}

The conventional picture of the turbulent cascade involves a specified outer \emph{injection} scale at which energy is supplied by large-scale processes (e.g., the MRI or large-scale torques) and specified smaller \emph{dissipation} scales at which the energy transforms into unordered kinetic motion (e.g., plasma kinetic scales or viscous/resistive scales in collisional systems). Solutions that bridge between these scales must conserve energy flux and are often assumed to only include interactions that are local in scale (e.g., only the eddies of similar sizes can efficiently interact with each other). The large separation between injection and dissipation scales often leads to assumption of ``zeroth law of turbulence,'' which states that the large-scale behavior of the cascade does not depend on the physics responsible for its dissipation. The assumption of scale-independence has been very useful in constructing models for collisional turbulence \citep{kolmogorov_1941_turbulence,goldreich_1995_alfvturb,boldyrev_2006_spectturb}. 

The RIAF systems most relevant for the EHT are much better described as Coulomb collisionless, but when the ratio of gas to magnetic pressure is large (which is a likely description of much of the accretion flow), perturbations in the magnetic field can drive sufficient deviations from local thermodynamic equilibrium to trigger kinetic micro-instabilities, which are non-local in nature and increase the effective collisionality beyond what the na\"{i}ve Coulomb collision picture implies \citep{kunz_2014_FirehoseMirrorInstabilities,sironi_2015_ElectronHeatingIon,Melville2016,Riquelme2016,Squire2017,Bott2021,Bott2023,Arzamasskiy2022,Ley2022,Tran2022}. This enhanced collisionality can lead to considerable dissipation close to the injection scale due to pressure-anisotropic viscous stress \citep{Arzamasskiy2022,Squire2023}. 

Several mechanisms for energy dissipation have been proposed in the context of turbulent cascades, including cyclotron heating \citep{KennelEngelmann1966,Isenberg2004,IsenbergVasquez2007,IsenbergVasquez2011,Arzamasskiy2019,Klein2020}, stochastic heating \citep{McChesney1987,Chandran2010,Arzamasskiy2019,Cerri2021}, Landau damping \citep{howes_2010_PrescriptionTurbulentHeating,kawazura_2020_IonElectronHeating,Arzamasskiy2022}, as well as reconnection and Fermi-type acceleration in relativistic plasmas \citep{zhdankin_2018_NumericalInvestigationKinetic,zhdankin_2019_ElectronIonEnergization,zhdankin_2021_ProductionPersistenceExtreme,Comisso2018,Comisso2019,Bacchini2022}. 

To avoid committing to a particular mechanism for dissipation, we adopt the sigmoidal $R_{\rm low}$--$R_{\rm high}$ model \citep{moscibrodzka_2016_rhigh,EHTC_2019_5}, in which the ion-to-electron heating ratio smoothly transitions between two asymptotic values in regions with low and high plasma $\beta = P_{\rm gas} / P_{\rm mag}$. Although this form is quite simplified, it is straight-forward to implement and the sigmoidal shape is qualitatively supported by some studies of energy dissipation in collisionless plasmas \citep[e.g.,][]{quataert_1998_ParticleHeatingAlfvenic,rowan_2019_ElectronProtonHeating,Arzamasskiy2022}. 

\subsection{The helicity barrier}
\label{sec:sigmac_definition}

When plasma $\beta \ll 1$, further complexities arise if the energies of waves propagating in opposite directions are unequal, i.e., when the turbulence is \emph{imbalanced}. This condition is typical in the case of the solar-wind plasma, but it may also occur in black hole accretion when strong outflows produce an imbalance biased along the outflow direction. How does imbalance alter the picture of a turbulent cascade? In imbalanced turbulence, the fluid is endowed with non-zero helicity, which must be conserved across the cascade in addition to the standard conservation of energy flux. In the inertial range (on scales $k_\perp \rho_{\rm i} \ll 1$ with $\rho_i$ the ion Larmor scale), both energy and helicity, which takes the form of a cross-helicity, can cascade simultaneously towards smaller scales; however, in the kinetic range, the cross-helicity is conservatively transformed into a magnetic helicity, the dispersion of the waves changes (Alfv\'en waves are converted into kinetic Alfv\'en waves, which are dispersive), and there is no solution that conserves the fluxes of both the energy and the generalized helicity.

The lack of solution results in an effective \emph{helicity barrier} \citep{Meyrand2021,Squire2022}, as the unbalanced portion of the cascading energy is trapped at scales $k_\perp \rho_{\rm i} \sim 1$. The energy accumulates at that scale until other cascade directions are enabled (\citealt{Squire2022} found that helicity barrier allows energy to enter a cascade of ion-cyclotron-waves, which eventually dissipate though ion-cyclotron heating). The imbalanced portion of the cascade thus only energizes the ions, and the maximum energy the electrons can receive is the balanced portion of the energy flux, which is itself divided between ions and electrons. The level of imbalance is quantified by the normalized quantity $\sigmac \in \left[-1, 1\right]$, whose absolute value increases to unity as the level of imbalance grows. The primary effect of the helicity barrier that we consider in this paper is thus the reduction of electron heating, $Q_{\rm e} \rightarrow (1-\sigmac) \, Q_{\rm e}$.

\vspace{0.5em}

To compute $\sigmac$, it is useful to work in the Els\"{a}sser formulation of magnetohydrodynamics. In the relativistic context and written in terms of the fluid four-velocity $u^\mu$ and magnetic field four-vector $b^\mu = - u_\nu \left(\star F\right)^{\mu\nu}$, with $\star F^{\mu\nu}$ the Hodge dual of the electromagnetic Faraday tensor, the Els\"{a}sser variables are \citep{chandran_2018_heatgrmhdturb}
\begin{align}
    z_{\pm}^\mu = u^\mu \pm \dfrac{b^\mu}{\sqrt{ \upvarepsilon}},
\end{align}
where the enthalpy is
\begin{align}
    \upvarepsilon = \rho + u + P + b^\alpha b_\alpha,
\end{align}
and where $\rho$ is the rest-mass density of the fluid, $u$ is its internal energy, and $P$ is its pressure. The standard interpretation of $z^\mu_\pm$ is that they describe the evolution of (pseudo-)Alfv\'{e}n waves propagating through the equilibrium magnetic field.

Describing the fluid as a mean background with fluctuations, the fluctuations are just the differences between $z_\pm^\mu$ and their locally time-averaged values,
\begin{align}
    \label{eq:deltazplus}
    \delta z^\mu_+ &= z^\mu_+ - \left< z^\mu_+ \right>, \\
    \delta z^\mu_- &= z^\mu_- - \left< z^\mu_- \right>,
    \label{eq:deltazminus}
\end{align}
and it is easy to show that the reduced relativistic Els\"{a}sser equations, written in terms of these difference variables, reduce to the standard equations of Newtonian reduced magnetohydrodynamics.

The Els\"{a}sser variables can be used to compute two ideal pseudoenergy invariants, $\left(\delta z^\mu_\pm\right)^2$, where we have introduced the shorthand $\left(v^\mu\right)^2 = v^\mu v_\mu$. The sum of the two pseudoenergies is the total energy in the system, and the difference measures the preference to generate waves in one direction or another (thus when the difference is non-zero, the system generates imbalanced turbulence). The normalized difference is the normalized cross-helicity:
\begin{align}
    \label{eq:unsigned_sigmac}
    \left|\sigmac\right| = \left| \dfrac{ \left(\delta z^\mu_+\right)^2 - \left(\delta z^\mu_-\right)^2 }{ \left(\delta z^\mu_+\right)^2 + \left(\delta z^\mu_-\right)^2 } \right| \, .
\end{align}

\vspace{1em}

Notice that there is ambiguity in how to perform the average in Equations~\ref{eq:deltazplus}~\&~\ref{eq:deltazminus}: When the system is variable, the fluid frame changes and the part of the electromagnetic field that is seen as the magnetic field by the fluid, $b^\mu$, changes with time. The quantities $\left< z^\mu_\pm \right>$ should represent the mean background flow; in regions where a characteristic background can be identified, performing a direct average of the four-vector components is then acceptable. In contrast, in regions that are highly variable, a mean background may not be readily identifiable and the meanings of $\delta z^\mu_\pm$ become less clear. In such regions, our averaging procedure yields smaller values for $\sigmac$, which one might heuristically expect since the rapidly varying magnetic field and flow geometry do not allow helicity to accumulate along a particular direction over a sustained period of time.
We compute $\left< z^\mu_\pm \right>$ as an average over the full duration of the simulations beginning after the transient from the initial conditions dies out. We have verified that decreasing the averaging window by a factor of two or four does not qualitatively change our results.

\vspace{0.5em}

\begin{figure*}[th!]
\centering
\includegraphics[width=\linewidth]{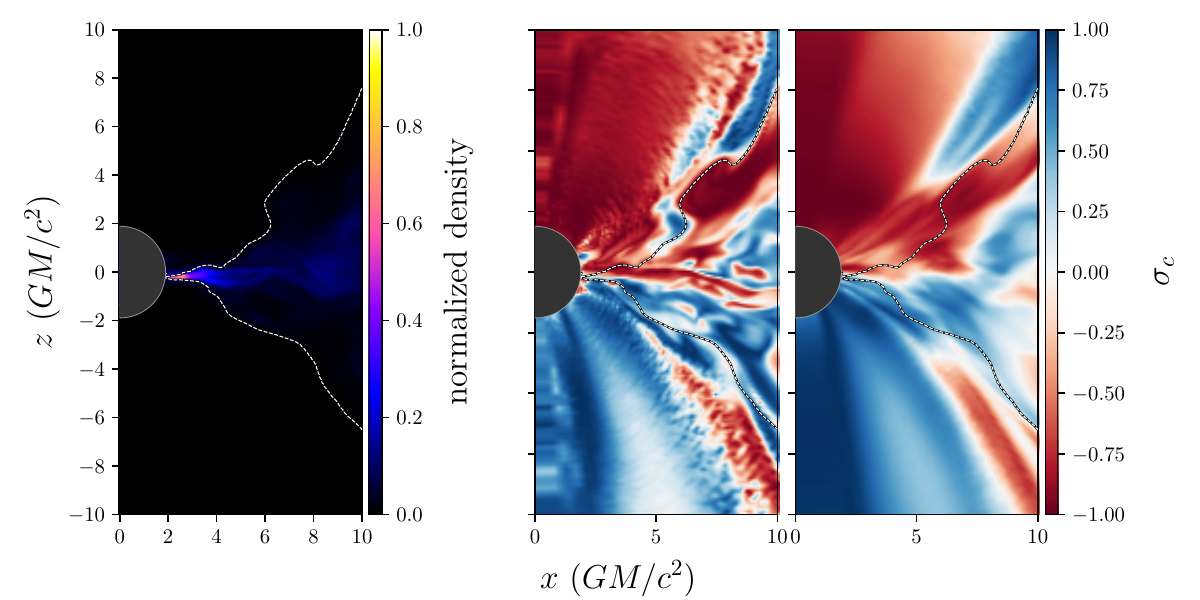}
\caption{
Effect of averaging signed $\sigmac$ over a dynamical time in an example snapshot from a MAD accretion flow with $\bhspin=0.5$. Left panel: normalized plasma density plotted in linear scale with contour showing the magnetization $\sigma = 1$ surface, within which all emission is produced. Right panels: signed $\sigmac$ across the domain with overplotted $\sigma$ contour. The center panel shows the instantaneously computed $\sigmac$ and the right-most panel shows $\sigmac$ averaged over one rotation period at $r = 3\,{\rm GM/c^3}$, where emission tends to peak. Although averaging smooths out smaller features, the overall magnitude of the final $\left|\sigmac\right|$ remains non-negligible. The grey circle at the left center of each panel outlines the black hole event horizon. 
}
\label{fig:average_vs_snapshot}
\end{figure*}

The value of $\sigmac$ in each fluid snapshot cannot be used directly to compute the effect of the helicity barrier, since the latter arises because of \emph{accumulated} cross-helicity. The physical picture is as follows: the cross-helicity injected at large-scales cascades down to smaller scales on the eddy turnover timescale, and eventually cross-helicity (of a particular sign) accumulates at the ion Larmor scale. The helicity barrier is effective only when cross-helicity has time to build up at the Larmor scale---injections of negative and positive cross-helicity at the large scales do cascade, but they ultimately cancel out. 

Because we perform our analysis in post-processing, we cannot track the cascade and injection of cross-helicity over time since that would require tracking the flow of non-zero cross-helicity fluid parcels as they evolve with the fluid. In this analysis, we instead approximate the buildup of cross-helicity by averaging the signed value of $\sigmac$ over approximately a dynamical time. This signed average is a good proxy for the total amount of accumulated cross-helicity in an axisymmetric flow with small radial velocities; we adopt this procedure even though our simulations are three-dimensional for the sake of computational efficiency. 

\vspace{0.5em}

Figure~\ref{fig:average_vs_snapshot} shows the signed value of $\sigmac$ in a snapshot compared against the average of $\sigmac$ over one dynamical time at $r = 3\,{\rm GM/c^3}$ where much of the observed emission is produced. The location of the disk can be identified by the plasma density, and the maximal extent of the emission region is bounded by magnetization $\sigma = 1$ contours. Evidently, the effect of averaging is to smooth out small-scale fluctuations in $\sigmac$ while the broader, large-scale features are left mostly unchanged. The sign of $\sigmac$ in the disk region is determined by the instantaneous flow properties. In MADs especially, cross-helicity of a particular sign may be long lived, as transient vertically asymmetric features are launched from large radii and fall through the event horizon. We discuss this smoothing procedure and compare between different averaging windows in the discussion section (see especially Figure~\ref{fig:beta2_averaging_examples}).

\section{Numerical methods}
\label{sec:numerical}

We use the {\tt{}PATOKA} pipeline to produce simulated images of RIAFs assuming the Kerr geometry~\citep{wong_2022_PATOKA}. The pipeline comprises a fluid simulation step, in which the general relativistic magnetohydrodynamics (GRMHD) code \iharm~\citep{gammie_2003_harm,prather_2021_iharm} produces the time evolution of the accretion flow in full 3D, and a ray-tracing step, in which the general relativistic radiative transfer code \ipole~\citep{moscibrodzka_2018_ipole} is used to compute the emission, extinction, and rotation of polarized light throughout the fluid simulation domain and track it to an observer at large distance.

\subsection{The fluid model}

The fluid evolution is obtained by solving the GRMHD equations, which take the form of a hyperbolic system of conservation laws
\begin{align}
\partial_t \left( \sqrt{-g} \rho u^t \right) &= -\partial_i \left( \sqrt{-g} \rho u^i \right), \label{eqn:massConservation}\\
    \partial_t \left( \sqrt{-g} {T^t}_{\nu} \right) &= - \partial_i \left( \sqrt{-g} {T^i}_{\nu} \right) + \sqrt{-g} {T^{\kappa}}_{\lambda} {\Gamma^{\lambda}}_{\nu\kappa},  \\
\partial_t \left( \sqrt{-g} B^i \right) &= - \partial_j \left[ \sqrt{-g} \left( b^j u^i - b^i u^j \right) \right], \label{eqn:fluxConservation}
\end{align}
along with the constraint
\begin{align}
\partial_i \left( \sqrt{-g} B^i \right) &= 0. \label{eqn:monopoleConstraint} 
\end{align}
Here, the plasma rest mass density is $\rho$ and its four-velocity is $u^\mu$. The magnetic field is represented by the $b^\mu$ four-vector. 
The spacetime geometry enters through the metric $g_{\mu\nu}$, its determinant $g$, and the Christoffel symbol ${\Gamma^{\alpha}}_{\beta\gamma}$. The symmetric rank-2 tensor $T^{\mu\nu}$ represents the stress--energy of the fluid, which has contributions from both the fluid and the electromagnetic field
\begin{align}
{T^{\mu\nu}} &= \left( \rho + u + P + b^{\lambda}b_{\lambda}\right)u^{\mu}u^{\nu} \nonumber  \\
&\qquad \quad + \left(P + \frac{b^{\lambda}b_{\lambda}}{2} \right){g^{\mu}}^{\nu} - b^{\mu}b^{\nu},
\label{eqn:mhdTensor}
\end{align}
where here $u$ is the internal energy of the fluid and the fluid pressure $P$ is related to its internal energy by a constant adiabatic index $\hat{\gamma}$ with $P = \left(\hat{\gamma} - 1\right) u$.

\subsection{Radiative transfer}

The time series fluid data are processed into simulated images with a radiative transfer post-processing step using the
\ipole code \citep{moscibrodzka_2018_ipole}. 
Each simulated image comprises a square grid of square pixels defined by a field-of-view (or width) in units of $GM/c^2$, distance from observer to source $d_{\mathrm{src}}$, and orientations with respect to the black hole spin axis and midplane (inclination and position angle). Pixels report the Stokes parameters $I_\nu, Q_\nu, U_\nu, V_\nu$ at their centers.

To construct an image, \ipole~first traces photon trajectories backward from the camera into the simulation domain by solving the geodesic equations
\begin{align}
    \dfrac{d x^\alpha}{d \lambda} &= k^\alpha \\
    \dfrac{dk^\alpha}{d\lambda} &= - {\Gamma^\alpha}_{\mu\nu} k^\mu k^\nu,
\end{align}
where $\Gamma$ is a Christoffel symbol, $\lambda$ is an affine parameter, and $k^\alpha$ is the photon wavevector. \ipole then integrates forward along each geodesic trajectory to solve the polarized radiative transfer equation, which in flat space can be written
\begin{align}
    { \scriptscriptstyle
    \label{eqn:polradxfer}
    \!\! \dfrac{d}{ds} \left( 
    \begin{matrix}
    I_\nu \\ Q_\nu \\ U_\nu \\ V_\nu
    \end{matrix}
    \right) = 
    \left(
    \begin{matrix}
    j_{\nu,I} \\ j_{\nu,Q} \\ j_{\nu,U} \\ j_{\nu,V}
    \end{matrix}
    \right) -
    \begingroup
    \setlength\arraycolsep{1pt}
    \begin{pmatrix}
    \alpha_{\nu,I} & \alpha_{\nu,Q} & \alpha_{\nu,U} & \alpha_{\nu,V} \\
    \alpha_{\nu,Q} & \alpha_{\nu,I} & \rho_{\nu,V} & - \rho_{\nu,U} \\
    \alpha_{\nu,U} & -\rho_{\nu,V} & \alpha_{\nu,I} & \rho_{\nu,Q} \\
    \alpha_{\nu,V} & \rho_{\nu,U} & -\rho_{\nu,Q} & \alpha_{\nu,I}
    \end{pmatrix}
    \endgroup
    \left( 
    \begin{matrix}
    I_\nu \\ Q_\nu \\ U_\nu \\ V_\nu
    \end{matrix}
    \right) },
\end{align}
where we have neglected scattering as its effect is negligible at the radio frequencies we are interested in. Here, emissivities $j_\nu$, absorptivities $\alpha_\nu$, and rotativities $\rho_\nu$ are frame-dependent quantities \citep{chandrasekhar_1960_RadiativeTransfer}. To compute the transfer coefficients, we use the thermal fits described in \citet{marszewski_2021_UpdatedTransferCoefficients}. Further detail about \ipole can be found in \citep{moscibrodzka_2018_ipole,wong_2022_PATOKA}. 

\vspace{0.5em}

Because GRMHD simulations introduce numerical floors in regions with high magnetization $\sigma = b^2 / \rho$, the plasma density and temperature are unreliable in such regions. To avoid contaminating the simulated images with numerical artifacts from the floors, we set the plasma density to zero in regions with $\sigma > 1$. Applying this $\sigma$-cutoff is reasonable, as the true density in highly magnetized regions like the jet is very small and therefore very little emission is produced there.

\subsection{Computing the electron temperature}
\label{sec:computing_electron_temperature}

Since the fluid simulations only track the total internal energy of the total fluid, there is freedom in assigning the electron distribution function. For M87*, radio frequency emission is produced by the synchrotron process \citep[e.g.,][]{yuan_2014_HotAccretionFlows}, and for the relevant plasma parameters, the $230\,$GHz emission observed by the EHT likely comes predominantly from the thermal core of the distribution function. We thus assume that the electron population can be modeled as a relativistic thermal Maxwell--J\"{u}ttner distribution, which is characterized by a single temperature $T_{\rm e}$.

The problem is thus to determine $T_{\rm e}$ given the total internal energy $u$ of the fluid and the local fluid properties, which requires partitioning the total internal energy $u$ into an ion component $\ui$ and an electron one $\ue$. Schematically, the internal energy can be written as
\begin{align}
    u &= \ui + \ue \nonumber \\
    &= \left(u_{\rm i,h} + u_{\rm i,z}\right) \quad + \quad \left(u_{\rm e,h} + u_{\rm e,z}\right) \nonumber \\
    &= \left( u_{\rm i,h} + u_{\rm e,h} \right) \quad + \quad \left( u_{\rm i,z} + u_{\rm e,z} \right) \nonumber \\
    &= u_{\rm h} + u_{\rm z}.
\end{align}
Here, we have used the subscript ${\rm h}$ (or ${\rm z}$) to denote the part of the internal energy that can be related to heating via the helical (or zero-helicity) part of the turbulent fluctuations. When $\beta$ is small, the helicity barrier stops any of $u_{\rm h}$ from cascading below the ion Larmor scale and heating the electrons, so $u_{\rm e,h} = 0$ subject to the condition that $\beta < \beta_{\rm critical}$. We set $\beta_{\rm critical} = 1$ to be consistent with the physical derivation of the barrier, but we have found that varying this cutoff value above unity has negligible impact on our results. 

\vspace{0.5em}

When $\beta < \beta_{\rm critical}$, we assume that $u_{\rm z} = \left(1 - \sigmac\right) u$, i.e., that energy imbalance is equal to injection imbalance. This equivalence is likely not true in general: \cite{schekochihin_2022_review} finds that energy imbalance is \emph{larger} than injection imbalance, although how well the quantitative details hold in non-idealized scenarios is uncertain. Nevertheless, under our assumption, the ion and electron energies are simply
\begin{align}
    \ui &= u_{\rm{i}, \rm z} + u_{\rm{i}, h} ,  \\
    \ue &= u_{\rm{e}, \rm z}.
\end{align}
For apples-to-apples comparison, we fix $R \equiv u_{\rm{i}, \rm z} / u_{\rm{e}, \rm z}$ regardless of $\sigmac$, which is reasonable under the approximation that the balanced component of the turbulent cascade is unaware of the imbalanced component. The ratio of total internal energies is then
\begin{align}
    R_{u} &\equiv \ui / \ue \\
    &= R_{u}(R, \sigmac) \\
    &= \dfrac{R + \sigmac}{1 - \sigmac}.
\end{align}

\vspace{0.5em}

Finally, to compute the electron temperature, we must find the relationship between the ion--electron temperature ratio $R_{T} \equiv \Ti / \Te$ and the energy ratio $R_{\rm u}$, which we do by assuming an ideal gas equation of state. Let the internal energies be
\begin{align}
    u_{\rm i} &= \left( \gamma_{\rm i} - 1 \right)^{-1} n_i \kB T_i, \\
    u_{\rm e} &= \left( \gamma_{\rm e} - 1 \right)^{-1} n_e kB T_e.
\end{align}
Taking $1/y$ and $1/z$ to be the number of electrons and nucleons (protons + neutrons) per (unionized) atom, respectively, then $n_{\rm e} = y \rho / m_p$, $n_{\rm i} = z \rho / m_p$, and we have that $n_{\rm i} = z n_{\rm e} / y$. The ratio of energies is therefore 
\begin{align}
R_{u} &= \dfrac{ \ui }{ \ue } \\
&= \dfrac{ \left( \gamma_{\rm e} - 1\right) }{\left(\gamma_{\rm i} - 1\right)} \dfrac{z/y \, n_e}{n_e} \dfrac{ \Ti }{\Te} \\
&= \dfrac{ z \left(\gamma_{\rm e} - 1 \right) }{ y \left( \gamma_{\rm i} - 1 \right) } R_{T} .
\end{align}
Assuming fully ionized hydrogen,\footnote{
Inferred brightness temperatures are in excess of $10^9\,$K, which is well above the ionization temperatures for both hydrogen and helium. The plasma composition is not well-constrained, however, and there may be non-trivial fractions of helium and heavier ions \citep{wong_2022_helium}.} which has $y = z = 1$, if the ions are nonrelativistic $\gamma_{\rm i} = 5/3$ and the electrons are relativistic $\gamma_{\rm e} = 4/3$, then
\begin{align}
    R_{T} = 2 R_{u} .
\end{align}

\vspace{0.5em}

We have not yet specified $R$, the ratio of ion-to-electron energies in the zero-helicity fluid component, which is in reality determined by the microphysics. To parameterize over this uncertainty, we let $R$ take any form allowed by the $R_{\rm low}$--$R_{\rm high}$ prescription \citep[see especially][]{moscibrodzka_2016_rhigh}
\begin{align}
    R = \dfrac{ \rlow + \rhigh \, \beta_{\mathrm{R}}^2 } { 1 + \beta_{\mathrm{R}}^2 },
    \label{eqn:rhigh}
\end{align}
which is motivated by models for electron heating in a turbulent collisionless plasma that preferentially heats the ions when the gas pressure exceeds the magnetic pressure.
Here $\beta_{\mathrm{R}} \equiv \beta / \beta_0$, and $\beta_0$, $\rlow$, and $\rhigh$ are parameters that control the temperature ratio in regions of low (high) $\beta$ where the plasma is dominated by gas (magnetic) pressure. The value of $\beta_0$ determines where the transition between $\rlow$ and $\rhigh$ occurs. We adopt typical values for $\rlow$ and $\beta_0$ and set them each to unity. Since plasma $\beta$ is large in disk regions, models with large $\rhigh$ mostly have cooler disks and, by contrast, hotter coron\ae{} and funnel walls, and thus often produce more emission from regions off the midplane.

\section{Results}
\label{sec:results}

We now use a set of GRMHD simulations to study the importance of the helicity barrier in simulated polarized observations of RIAFs. For simplicity, we focus on the M87* accretion system and so set the black hole mass to $M = 6.5 \times 10^9\, M_\odot$ for consistency with observational results (see Table 1 of \citealt{EHTC_2019_6}). This mass choice provides a physical length scale to the simulations. We use Equation~\ref{eq:unsigned_sigmac} and the averaging procedure described above to compute $\sigmac$ across the domain and calculate electron temperatures. We image each snapshot of the fluid simulation twice, once using electron temperatures computed incorporating the helicity barrier and once with the effects of the helicity barrier turned off. For each time series of images, we rescale the mass density of the accreting plasma until the average of the $230\,$GHz flux density light curve matches the observed (instantaneous) value of $F_{230\,\mathrm{GHz}} = 0.65\,$Jy (see Appendix B.1 of \citealt{EHTC_2019_4}). See Appendix D of \citealt{wong_2022_PATOKA} for caveats and more detail about the flux-fitting procedure.

To compare against observations, the camera must be assigned both an inclination and a position angle, which we define relative to the jet launched by the system. For our target M87*, there is clear evidence of a large-scale jet  (see \citealt{walker_2018_m87jet}) with a measured inclination angle of $17^\circ$ relative to the line of sight. We therefore orient our camera at either $17^\circ$ or $163^\circ$ relative to the axis of the jet in the simulation (this is coincident with the black hole spin axis) according to which parity reproduces the observed brightness asymmetry seen in \citet{EHTC_2019_5}, which manifests as a greater brightness temperature in the bottom half of the image. The statistical axisymmetry of the accretion flow means that rotating the image to align the position angle of the jet with its observed value does not influence the other images statistics. We thus fix the position angle of the images so that the jet lies in the vertical direction, as determined by the default $\mathrm{PA} = 0$ setting for the simulations.

\subsection{Model parameter space}

\begin{figure*}[th!]
\centering
\includegraphics[width=\linewidth]{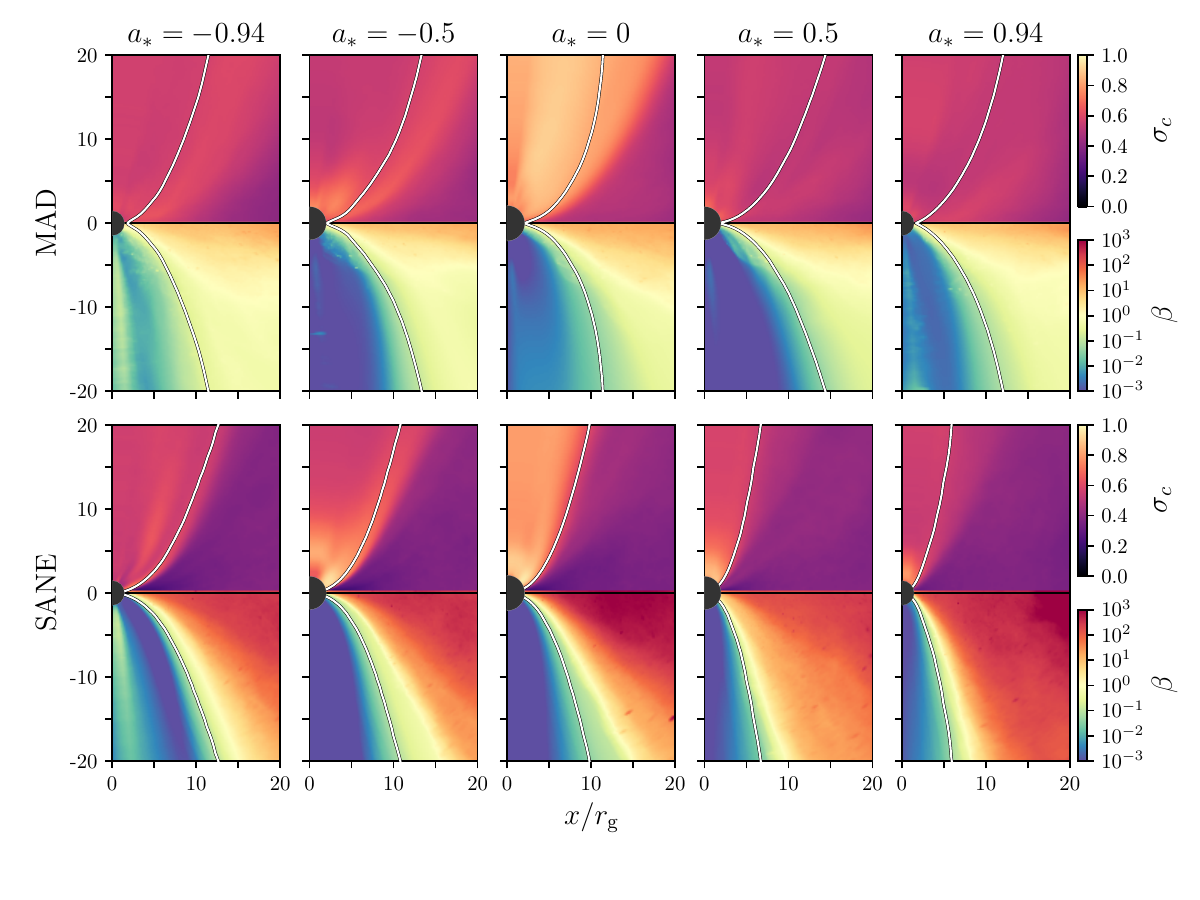}
\caption{
Average value of $\left|\sigmac\right|$ and plasma $\beta$ across domain for ten different GRMHD simulations of RIAFs across MAD and SANE states and five different black hole spins $\bhspin = -0.94, -0.5, 0, 0.5$, and $0.94$. The average for $\left|\sigmac\right|$ is computed from the value of $\left|\sigmac\right|$ calculated for each snapshot according to the signed averaging procedure described in \S\ref{sec:sigmac_definition}. The accretion flow is concentrated in disks near the midplane, where $\beta$ is largest. The magnetization $\sigma = 1$ surface is plotted as a white line in each panel; $\sigma$ increases towards the poles and decreases in the midplane. In our models, the vast majority of emission is produced in regions with $\sigma < 1$, so from an observational perspective the value of $\sigmac$ between the two white lines matters the most.
}
\label{fig:sigmac_models}
\end{figure*}

The space of possible accretion configurations is high dimensional, covering the black hole mass and angular momentum parameters, the accretion rate of the system, boundary conditions and gas composition, and the magnetic field configuration. It is not computationally feasible to explore the full parameter space, so we focus on the subset corresponding to the canonical models used in the initial Event Horizon Telescope analysis of M87*. We thus aim to identify general trends and gauge the overall importance of the helicity barrier rather than make quantitatively precise predictions.

\vspace{0.5em}

The magnetization of an accretion system can be used to differentiate flows according to whether the magnetic pressure near the horizon is strong enough to counterbalance the inward ram pressure of the fluid. When the magnetic pressure is high enough, the infalling motion of the plasma is arrested and the flow enters the magnetically arrested disk (MAD; \citealt{bisnovatyi-kogan_1974_AccretionMatterMAD,igumenshchev_2003_ThreedimensionalMagnetohydrodynamicSimulations,narayan_2003_MagneticallyArrestedDisk}) state. The alternative scenario is canonically referred to as standard and normal evolution (SANE; \citealt{narayan_2012_sane,sadowski_2013_EnergyMomentumMass}). SANE flows are turbulent but steady; in MAD flows, large tubes of magnetic flux arrest the inward motion of the flow and accretion is chaotic and mediated by transient filaments of hot plasma that thread the region between the hole and plasma at large radius. We consider both the MAD and SANE accretion states.

\begin{figure*}[th!]
\centering
\includegraphics[width=\linewidth]{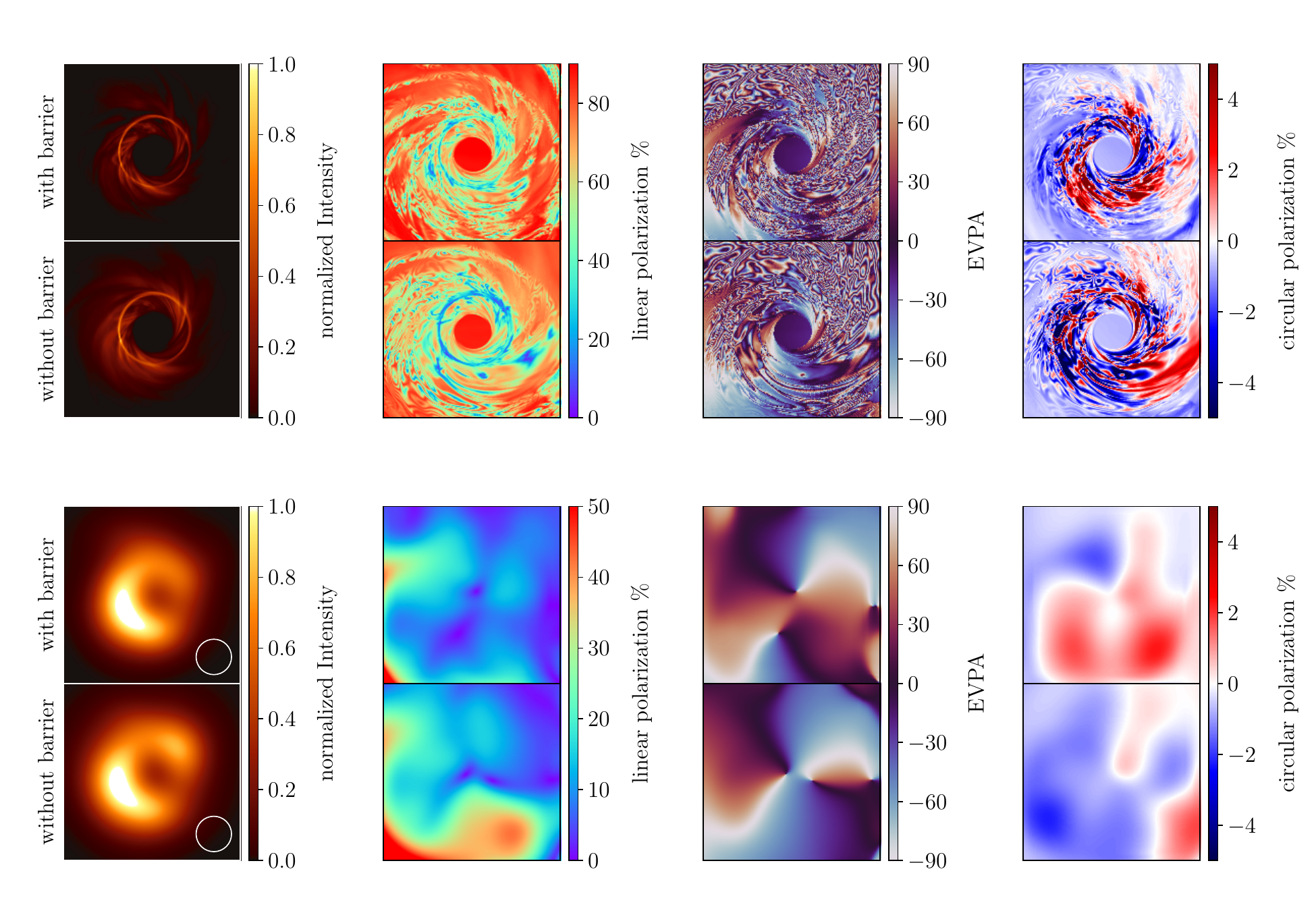}
\caption{
Example polarimetric image from a MAD $\bhspin = 0.5$ $\rhigh=40$ model snapshot with and without including the effect of the helicity barrier on the electron heating model. Columns show the total intensity, linear polarization $\sqrt{Q^2 + U^2}/I$, electric vector position angle $\frac{1}{2} \arctan U/Q$ (measured east-of-north or counterclockwise-from-vertical on the sky), and the degree of circular polarization $V/I$, respectively. The bottom two rows show the images after being blurred with a $20\,\mu{\rm as}$ Gaussian beam to simulate the effective resolution of the Event Horizon Telescope. Since including the helicity barrier produces cooler electrons, the number density of the plasma must be increased to compensate, leading to more depolarization in the polarimetric signature (especially visible in the lower parts of the EVPA images). It also leads to less lensed emission coming through the disk in the photon ring, which produces less depolarization. While the detailed structure of the circular polarization can change significantly, the overall level of circular polarization does not tend to change.
}
\label{fig:example_image_comparison}
\end{figure*}

We express the black hole angular momentum in terms of the dimensionless spin parameter $\bhspin \equiv J c / GM^2$ with $\left| \bhspin \right| \le 1$, where $J$ is the magnitude of the angular momentum. By convention, we set $\bhspin < 0$ when the angular momentum of the accretion flow and the spin of the black hole are anti-aligned. There is no reason that the angular momenta of the hole and the flow must be precisely aligned or anti-aligned. Tilted systems have recently gained broad attention; for simplicity, however, we restrict our focus to systems with no tilt. We consider five black hole spins $\bhspin = -15/16, -1/2, 0, 1/2,$ and $15/16$ (hereafter written as $-0.94, -0.5, 0, 0.5, 0.94$ to be consistent with EHTC publications).

\vspace{0.5em}

Although computing the radiative transfer coefficients requires choosing mass-density and length scales, since the GRMHD equations and Equation~\ref{eq:unsigned_sigmac} are invariant under these rescalings, it is possible to measure the degree of cross-helicity directly from the scale-free fluid snapshot variables before restricting to a particular observer inclination or black hole accretion system. In Figure~\ref{fig:sigmac_models} we show the simulation-averaged values both for $\left|\sigmac\right|$ and plasma $\beta$. For $\left|\sigmac\right|$, we have computed the time-average of the absolute value of the signed quantity $\sigmac$ that has been calculated per fluid snapshot as described in Section~\ref{sec:sigmac_definition} (and shown in the right-most panel of Figure~\ref{fig:average_vs_snapshot}). We show the average of the absolute value to account for the fact that the infall timescale is often shorter than the timescale over which $\sigmac$ changes sign, since the magnitude of $\sigmac$ controls the helicity barrier. Thus, the non-zero imbalance in the midplane of the MADs is due to spontaneous symmetry breaking that does not average out before the fluid parcels carrying the cross-helicity fall through the event horizon. Different choices for averaging windows are considered in the discussion (see especially Figure~\ref{fig:beta2_averaging_examples}).

In regions where $\beta$ is large, the accretion flow takes the form of a turbulent disk, fluctuations are large, and $\sigmac$ is smallest. This effect is most prominently seen in the SANE flows and flows with small $\bhspin$. Since MAD flows have more consistent magnetic fields, $\sigmac$ keeps the same sign over longer timescales; this is reflected in the characteristically larger values of $\sigmac$ in the MAD flows. 

\begin{figure*}[ht!]
\centering
\includegraphics[width=0.85\linewidth]{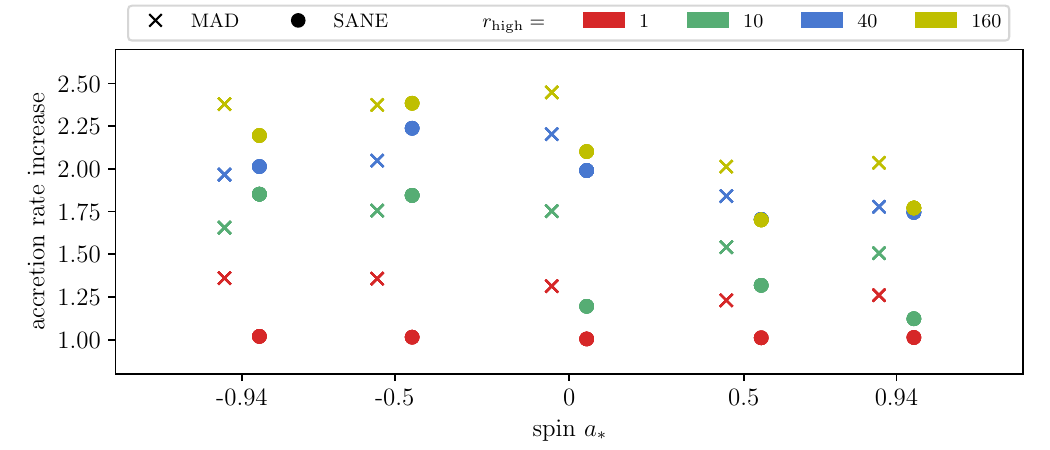}
\caption{
Relative increase in accretion rate $\dot{m}$ for all library models after including effect of the helicity barrier. Colors correspond to different values of $\rhigh$. The helicity barrier results in cooler electrons, which require higher local number densities to produce the same observed target flux. The higher number densities correspond to larger accretion rates. Systems with large $\rhigh$ tend to exhibit the greatest increase, as the effect of the helicity barrier is highest in the jet regions, where much of the emission is produced in large-$\rhigh$ models. MAD models typically have larger overall increases in $\dot{m}$, since $\sigmac$ is typically larger across their entire domains (see Figure~\ref{fig:sigmac_models}). The spin of each model is given by the closest labeled value on the $x$-axis.
}
\label{fig:accretion_rate_difference}
\end{figure*}

In all cases, the helicity barrier operates most strongly in regions with the most ordered magnetic field. In SANE flows the most ordered fields live within the jet and its enveloping wind. These regions have low plasma $\beta$ and are approximately bounded by the magnetization $\sigma = 1$ contour.
In MAD flows, the field maintains order throughout the domain and helicity builds up nearly equally everywhere. The ``funnel'' regions in the low-spin cases exhibit particularly ordered fields, which arise as accreted magnetic field lines build up near the horizon and are less perturbed by, e.g., the strong torques that a spinning black hole would impart on them due to frame dragging.

\subsection{Images and emission source}

We have thus far explored $\sigmac$ in global accretion models from an observer-agnostic perspective. To understand how the helicity barrier influences observables, it is necessary to adopt an emission model, i.e., we must both choose thermodynamic flow parameters and set the observer inclination.
We will focus on observations of the M87* accretion flow targeted by the EHT.

We use the \ipole code to produce polarimetric images at the $230\,{\mathrm{GHz}}$ operational frequency of the EHT. In Figure~\ref{fig:example_image_comparison}, we show example images produced from the same single fluid snapshot shown in Figure~\ref{fig:average_vs_snapshot} evaluated using thermodynamic models that either do or do not incorporate the influence of $\sigmac$ on the ion--electron energy partition as described in Section~\ref{sec:computing_electron_temperature}. Columns show the full polarized properties of the light, including total intensity, degree of local linear polarization $\sqrt{Q^2 + U^2}/I$, electric vector position angle $\frac{1}{2} \arctan U/Q$ (measured east-of-north or counterclockwise-from-vertical on the sky), and the degree of circular polarization $V/I$, respectively. The bottom two rows of Figure~\ref{fig:example_image_comparison} show the same images at the top two rows but blurred with a $20\,\mu{\rm as}$ Gaussian to simulate the effective resolution of the Event Horizon Telescope. This blurring is particularly important when considering observations of, e.g., resolved linear polarization, which may be high when the resolution element is smaller than the spatial correlation length of the EVPA but which is decreased dramatically when blurring over regions with a rapidly varying EVPA. 

\begin{figure*}[th!]
\centering
\includegraphics[width=0.7 \linewidth]{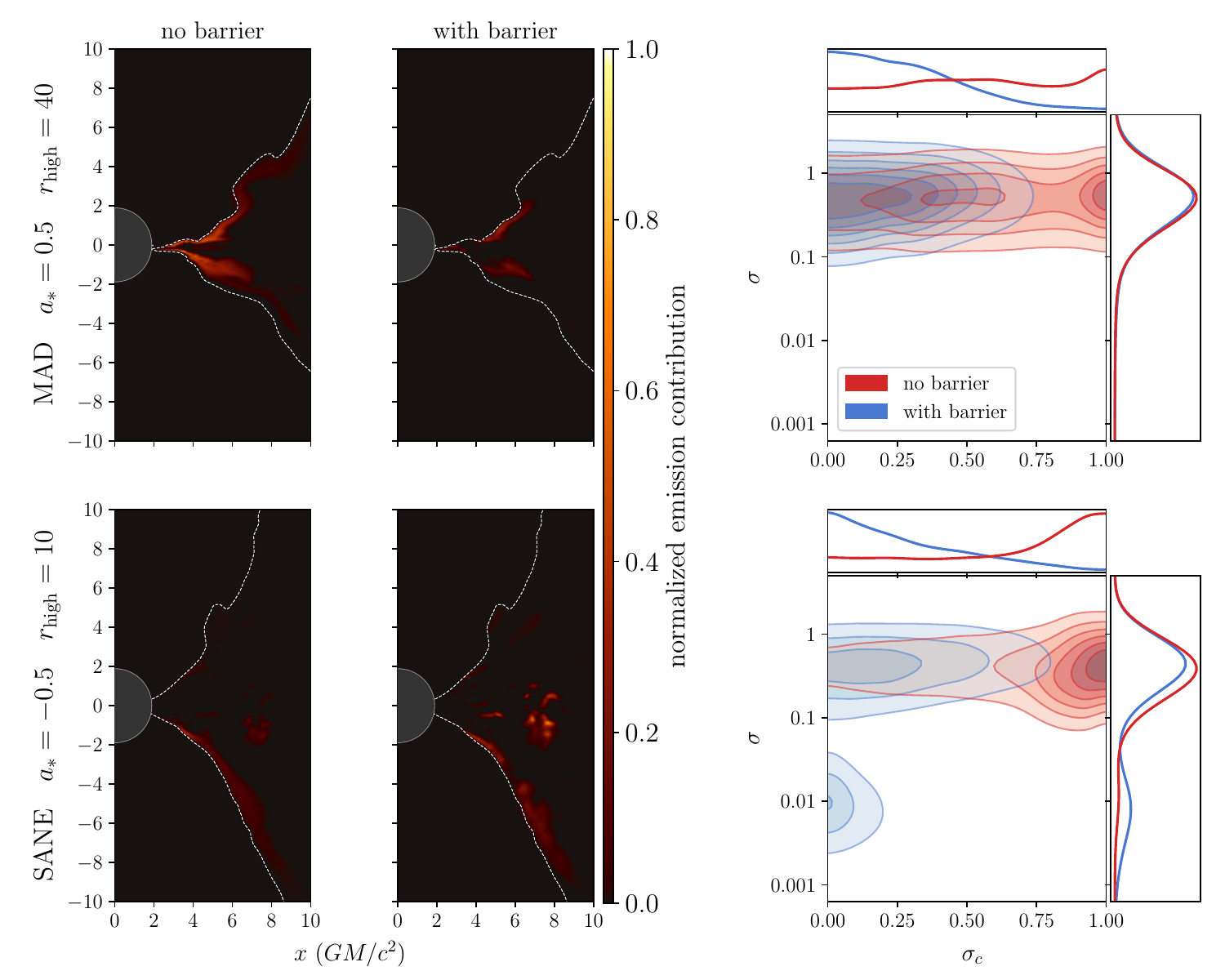}
\caption{
Location and plasma properties of emitting regions for two sample GRMHD simulation snapshots. Only emission that contributes to the final image is considered (see~\citealt{wong_2022_PATOKA} for more detail). The left panels show the effect of the helicity barrier on the geometry of the emission region. The right panels show histograms of where the emission originates by $\sigmac$ and magnetization $\sigma$. In MADs the value of $\sigmac$ is comparable across the domain so that emission everywhere is decreased but significantly change in morphology; in SANEs, $\sigmac$ is greater along the funnel walls and thus the relative contribution from near the funnel vs.~the midplane decreases. The white dashed lines show the magnetization $\sigma=1$ contour, outside of which emission has been disallowed. 
The accretion rate for each of the models has been adjusted so that the flux density at $230\,$GHz for each image is $0.65\,$Jy to be consistent with observations of M87*.
}
\label{fig:emission_source_comparison}
\end{figure*}

Although the images with and without the barrier are produced from the same fluid model, they correspond to \emph{different} accretion rates selected such that the average flux density is $0.65\,{\mathrm{Jy}}$ to be consistent with observations. Thus, although the morphology of the fluid in the underlying accretion flows is the same for the different images, the number density and magnetic field strength differ. Figure~\ref{fig:accretion_rate_difference} shows the factor by which the accretion rate must be increased for the flux to match observations. For internal consistency with the scale-free GRMHD equations, the increased accretion rate requires any local energy density quantity  be increased by the same factor. In our case, the plasma number density, the fluid internal energy, and the square of the magnetic field strength must all be increased by the value shown in Figure~\ref{fig:accretion_rate_difference}. MADs and especially large $\rhigh$ models have the largest required increase, as emission in those systems tends to be in regions with the largest imbalance and the greatest importance of the helicity barrier. Incorporating the helicity barrier yields higher estimates for jet power, since the jet power scales directly with the accretion rate.\footnote{The relative power of the jet compared to the infalling rest-mass energy is determined by the simulation. When the plasma number density is increased to match the observed flux, all energy densities must be rescaled by the same factor, so the absolute jet power scales with the accretion rate.}

\vspace{1em}

\begin{figure*}[th!]
\centering
\includegraphics[width=\linewidth]{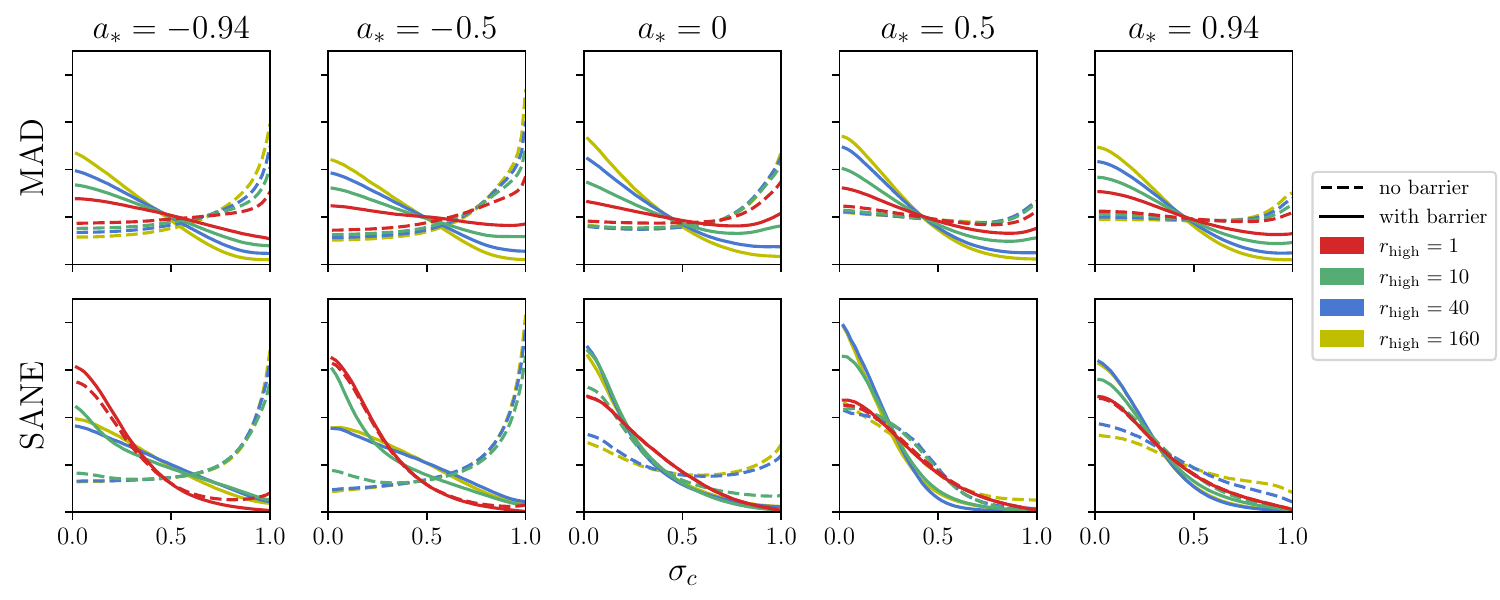}
\caption{
Comparison of emission-weighted $\sigmac$ for library models with (solid) or without (dashed) the effect of the helicity barrier. The area under each curve has been normalized to unity. MADs have larger values of $\sigmac$ across their entire domains, and the influence of the helicity barrier is evident in the change of the curve shape for all models. In SANE models with low $\rhigh$, emission primarily comes from the disk region, which already has low values of $\sigmac$. Thus, the helicity barrier has little impact on the SANE $\rhigh = 1$ models. As $\rhigh$ is increased in the SANEs, more emission comes from the funnel wall region and the effects of the helicity barrier's operation become more evident.
}
\label{fig:emission_source_overview_histograms}
\end{figure*}

Given an emission model, it is possible to evaluate how the helicity barrier alters the source morphology. In Figure~\ref{fig:emission_source_comparison}, we show the location of the emission in both the MAD model of Figures~\ref{fig:average_vs_snapshot} \& \ref{fig:example_image_comparison} as well as a representative SANE model. The MAD and SANE models have different spins and different electron thermodynamics. The right panels of the figure show the characteristic magnetization $\sigma$ and cross-helicity $\sigmac$ of the emission. The funnel region that lies at the interface between the jet core and the disk typically has larger values of $\sigma$ than the disk. 

As expected, including the effects of the helicity barrier limits emission from regions with large $\sigmac$. Figure~\ref{fig:sigmac_models} shows that emission in MAD models does not change drastically while in SANE models the emission tends to shift away from the funnel wall and toward the lower-magnetization disk region. The right panels of Figure~\ref{fig:emission_source_comparison} show this trend as well: emission shifts from regions of high $\sigmac$ to small $\sigmac$ while the characteristic magnetization $\sigma$ in the emission regions shifts from large values to small values in the SANE flow.

Figure~\ref{fig:emission_source_overview_histograms} shows how the emission source changes across all library models. Emission in regions with large $\sigmac$ decreases as expected. MADs typically produce emission throughout their infalling regions regardless of the thermodynamics prescription; as more of their domain has large values for $\sigmac$, the effect of the helicity barrier is very evident as emission in regions with large $\sigmac$ drops significantly, altering the shape of each curve.
SANE models with significant funnel-wall emission (i.e., models with large values of $\rhigh$) are often the most strongly affected. SANE models with low $\rhigh$, i.e., models where the majority of the emission comes from the disk, are almost completely unaffected. 

\subsection{Polarization}

\begin{figure*}[h!]
\centering
\includegraphics[width=\linewidth]{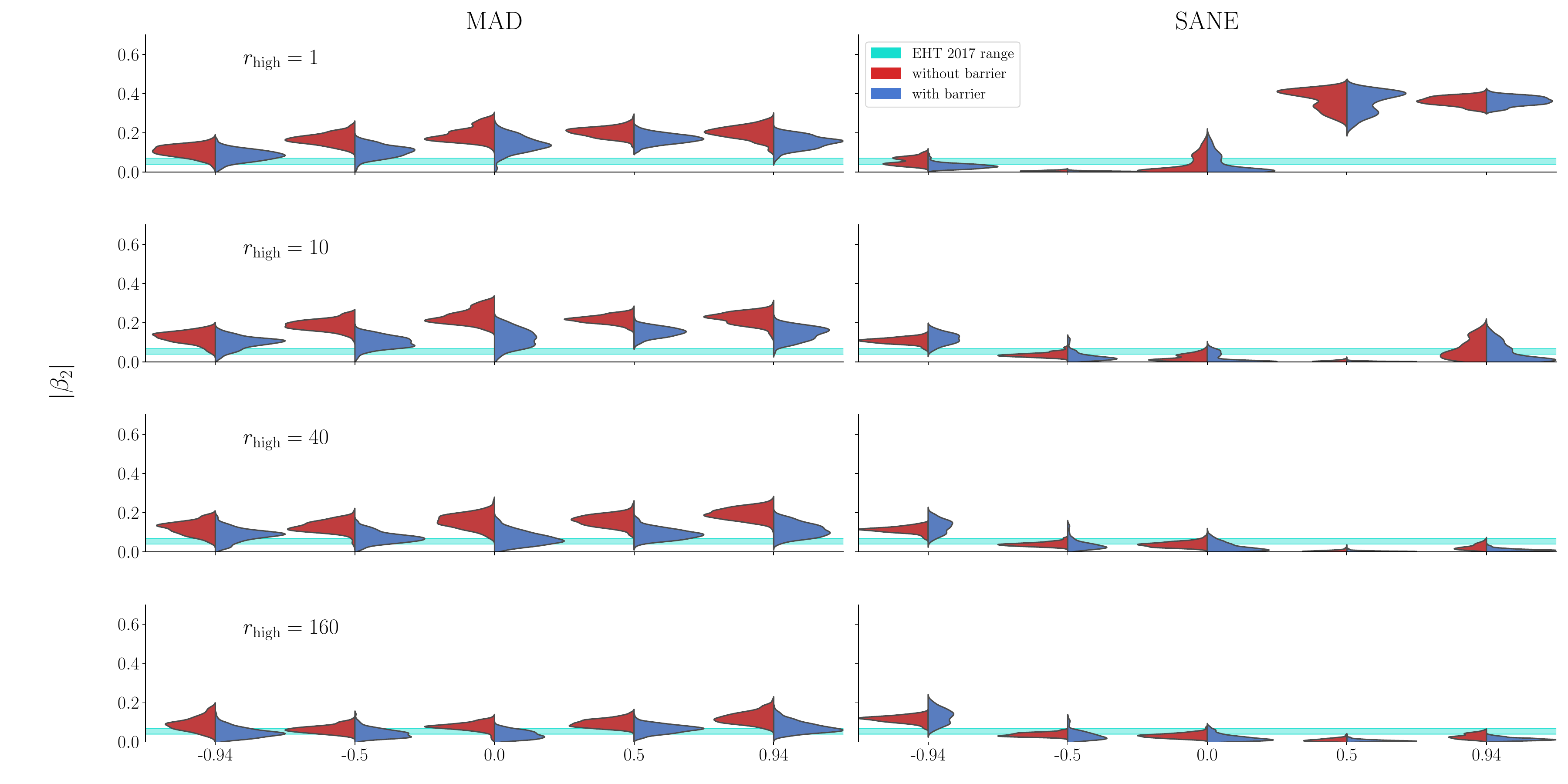} \\
\includegraphics[width=\linewidth]{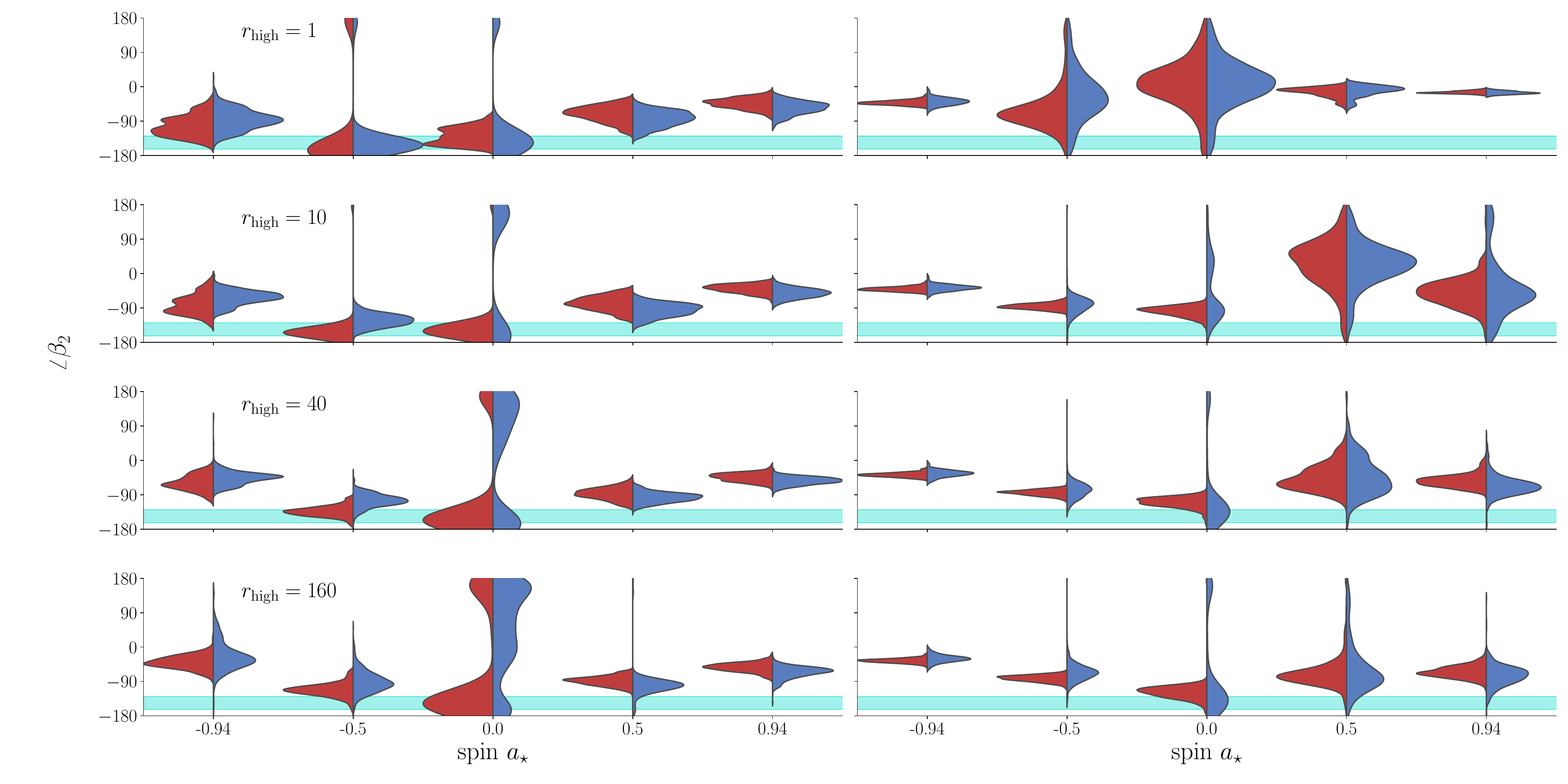} 
\caption{
The effect of including the helicity barrier in the electron temperature model on the polarimetric observable $\beta_2$ across the library models. The amplitude of the $\beta_2$ coefficient tends to decrease more in MAD models due to increased scrambling across the image from higher Faraday depths. In all models, the distribution of $\angle \beta_2$ is less affected, although sometimes tends towards the more toroidal configuration. The trend of $\angle \beta_2$ vs.~$\bhspin$ remains unchanged.
}
\label{fig:beta2_overview_allmodels}
\end{figure*}

We now evaluate how including the effect of the helicity barrier can affect the linear polarimetric $\beta_2$ observable, which has been used by the EHT to gauge the strength of the horizon-scale magnetic field and differentiate between different accretion models \citep{palumbo_2020_DiscriminatingAccretionStates,EHTC_2021_8}. The complex $\beta_2$ coefficient measures the power in (amplitude) and orientation of (argument) the azimuthally symmetric mode of the linear polarization vector across the image. The final value of the $\beta_2$ coefficient is determined by both the structure of the magnetic field in the emitting regions of the flow and the degree of depolarization due to differences in, e.g., Faraday rotation as the light propagates through the flow. Since the spin of the black hole influences the structure of the magnetic field, there is a trend of $\angle \beta_2$ with spin, with higher values of $\left| \bhspin \right|$ producing more toroidal fields and pushing $\angle \beta_2$ towards zero (radial linear polarization pattern). It is worthwhile to understand how plasma physics uncertainties might complicate this relationship.

\begin{figure*}[ht!]
\centering
\includegraphics[width=0.8\linewidth]{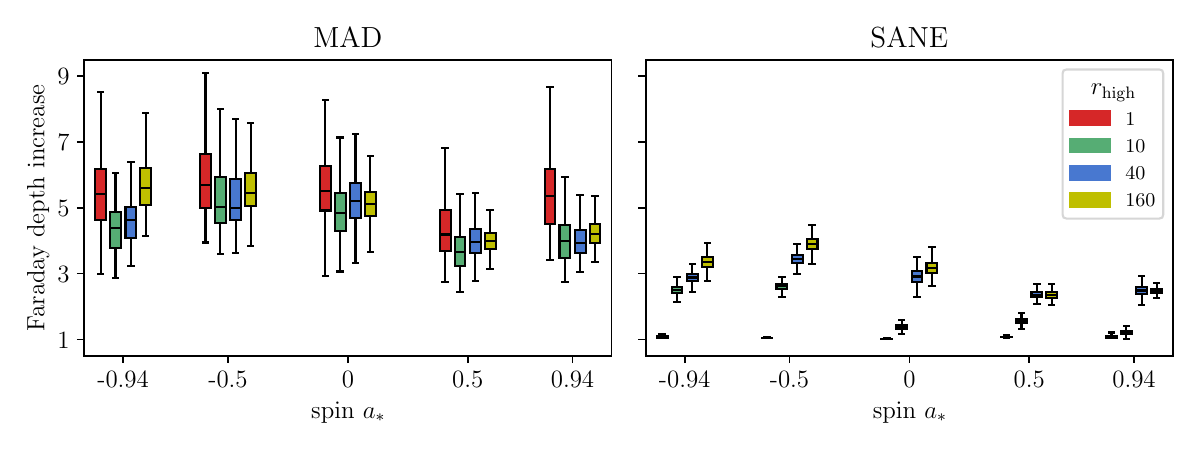}
\caption{
Approximate increase in Faraday depth across all images from all models in our library when the helicity barrier is turned on. We compute the Faraday depth for each image as a polarization-weighted average over all image pixels of the Faraday depth along the entirety of each geodesic. Since MAD flows see large $\sigmac$ across their full domains, the plasma density in emitting regions increases significantly and is reflected in the increased Faraday depth. The effects of the helicity barrier in SANEs is more confined to the jet funnel region, and so the increase in Faraday depth is largest when emission comes from those regions, i.e., especially in the models with large $\rhigh$. The box extends between the first and third quartiles and the whiskers extend the standard $1.5\,\times\,$the interquartile range.
}
\label{fig:faraday_overview}
\end{figure*}

\begin{figure*}[ht!]
\centering
\includegraphics[width=0.8\linewidth]{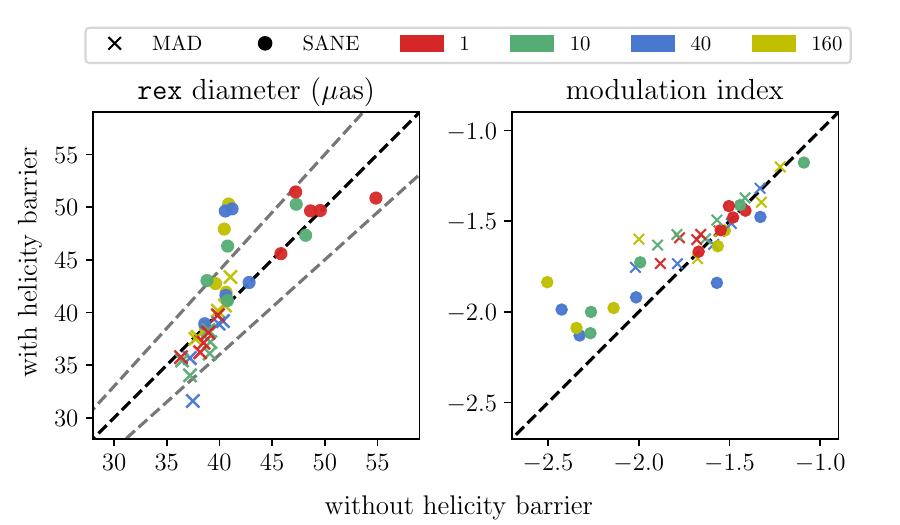}
\caption{
Exploration of how including the effect of the helicity barrier on electron temperatures changes {\tt{rex}}-fit diameters (left) and variability (modulation index; right) across library models. The color of each point indicates the model $r_{\rm high}$. Points would lie on the black lines if the helicity barrier had no effect; the grey dashed lines in the {\tt{rex}} panel show $\pm 10\%$ deviation. Points in the lower-right triangle are smaller or less variable with the helicity barrier. For both statistics, MAD models are mostly unaffected since their emission regions remain similar. SANE models with larger $\rhigh$ tend to shift emission towards the disk when the helicity barrier is included, which results in more extended emission and larger apparent ring diameters. The variability for these models increases since it is controlled by the turbulent dynamics of the infalling plasma closer to the event horizon instead of in the funnel wall (notice that the points trend towards the value of the red $\rhigh=1$ SANE points).
}
\label{fig:rex_variability}
\end{figure*}

Figure~\ref{fig:beta2_overview_allmodels} shows how both the amplitude and argument of $\beta_2$ change for the different models in our library. Broadly, the amplitude of $\beta_2$ decreases with the inclusion of the helicity barrier while the argument of $\beta_2$ is mostly unaffected. The amplitude typically decreases the most in MAD models. Since $\sigmac$ in MAD flows is mostly consistent across the domain, the regions that contribute to the image are mostly unchanged so that the general image structure persists. The differences are instead due to the reduced emission-per-particle due to lower temperatures, which must be compensated by an increased number density. This renormalization results in both an increased accretion rate but, more importantly, also increased depolarization since the differences in the increased column density of plasma along neighboring lines of sight lead to more extreme differing levels of rotation over the course of the light's propagation.

The differences that produce more scrambled images are quantitatively related to the Faraday depth along the geodesics as emission travels from its source to the observer. Figure~\ref{fig:faraday_overview} shows a proxy for the factor by which the Faraday depth increases when the effects of the helicity barrier are included. Our proxy is computed by first evaluating the total Faraday depth along the full geodesic for each pixel in the image and then computing the polarization $P = \sqrt{Q^2 + U^2}$-weighted average of these values over all image pixels. The increase in Faraday depth is significant in MAD flows for all models; in SANE flows, the Faraday depth increases are more evident in models with large $\rhigh$, where the emission is more likely to arise in the large-$\sigmac$ jet funnel regions.

This scrambling effect can be seen in the EVPA panels of Figure~\ref{fig:example_image_comparison}, especially in the lower parts of the images where the bottom panel (with its lower density) has more coherent EVPA compared to the top panel). In the blurred linear polarimetric maps in the same figure, it is clear that the linear polarization in the bottom-right (southwest) part of the image decreases because the overall intensity is canceled out by the near-random phases of the neighboring pixels' EVPAs. In SANE models the polarization pattern is often already highly scrambled because the magnetic fields in the emission region are highly disordered. Since the images start out scrambled, increasing the number density of the flow does not have as noticeable an effect.

The increased optical depth through the disk also means that the image of the lensed photon ring will appear \emph{less} depolarized (see the blue ring that appears in the ``without barrier'' resolved linear polarization image of Figure~\ref{fig:example_image_comparison}). When the disk is optically thin, each pixel contains contributions from the direct image as well as the lensed secondary (and so on) images. The lensed images exhibit a conjugate polarization signature; the contributions from the lensed images cancel in part and the summed final polarization signal is decreased (for more detail see \citealt{himwich_2020_UniversalPolarimetricSignatures,palumbo_2022_photonringbeta2}). The increased column density due to the effect of the helicity barrier on the temperatures means that the secondary image is less prominent and thus less cancellation happens along the relevant trajectories.

\subsection{Ring diameter \& variability}

Finally, we check whether disregarding the helicity barrier can bias several other parameters inferred by the EHT. Here, we focus on the ring diameter \citep{EHTC_2019_6,psaltis_2020_gravtesteht,EHTC_2022_5,EHTC_2022_6}, which has been used to test consistency of the observational data with the theory of general relativity, and on the variability in the compact-flux light curve, which has demonstrated notable disagreement between models and the observational data \cite{EHTC_2022_5}. 

We measure a ring diameter for each image in our library with the ring extractor (\texttt{rex}) method described in \S9 of \citet[][see also \citealt{chael_2022_ehtim}]{EHTC_2019_4}. \texttt{rex} makes its measurement from the algorithmically identified ``center point'' of each image---the point that is most equidistant from the peak intensity along each of 360 equally-spaced rays cast from itself. The left panel of Figure~\ref{fig:rex_variability} shows a sampling of ring diameters taken from our M87*-like models, which are at low inclination where a ring diameter measurement is easiest to perform. As can be seen in the figure, disregarding the helicity barrier in MAD models tends to increase the measured ring diameter slightly although the overall measurements stay roughly consistent. In SANE models the helicity barrier alters the electron temperatures such that the measured ring diameter is roughly consistent or slightly larger. The largest increase in measured ring diameter occurs for the models with large $\rhigh$, where emission in the jet funnel is suppressed and the disk contributes much more significantly to the image. 

\begin{figure}[th!]
\centering
\includegraphics[width=\linewidth]{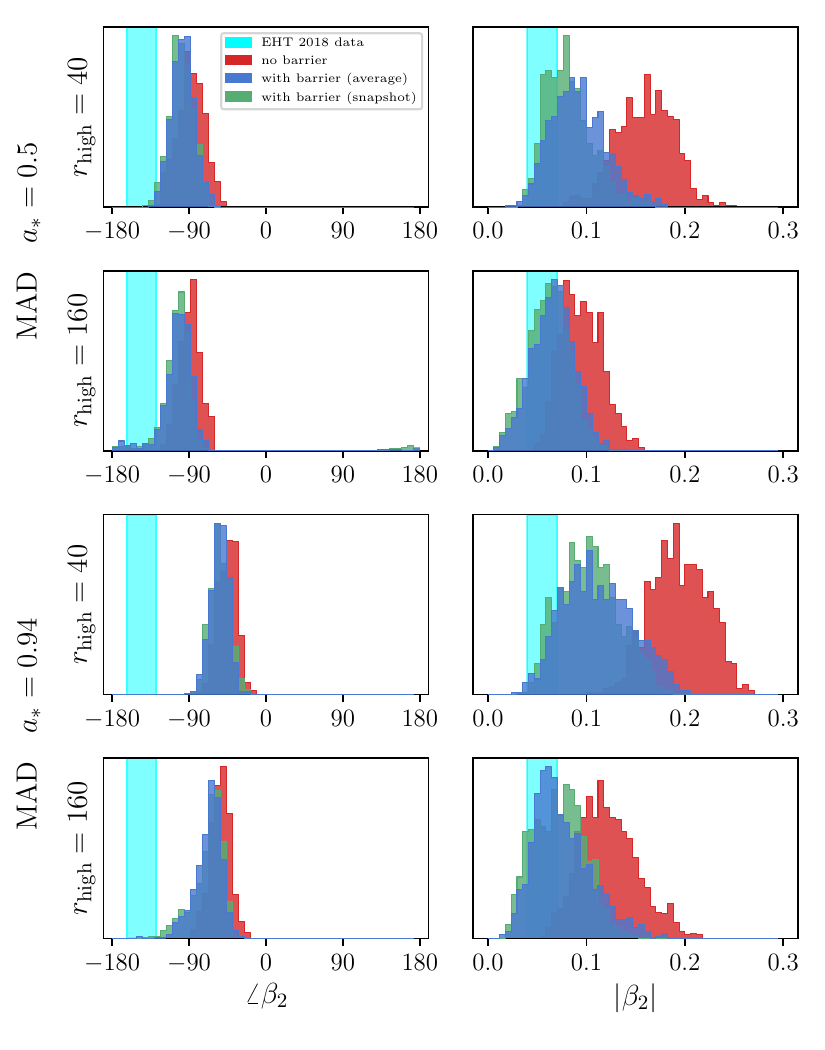}
\caption{Comparison of measured $\beta_2$ values from representative accretion models for two different choices for the averaging windows of the signed $\sigmac$ quantity. The blue ranges corresponds to an average over one dynamical time at the characteristic emission radius $r = 3\,{\mathrm{GM/c}}^3$ and is the choice adopted in the rest of the paper. The green ranges show the effect of removing this average, which tends to increase the absolute value of $\sigmac$. The red ranges show the values measured before incorporating the helicity barrier. The cyan bar corresponds the values inferred from EHT observations of M87* \citep{EHTC_2021_7}.
}
\label{fig:beta2_averaging_examples}
\end{figure}

The right panel of Figure~\ref{fig:rex_variability} shows how the measured modulation index changes when the effect of the helicity barrier is included in the electron thermodynamics calculation. The modulation index is
\begin{align}
    M_{\Delta T} \equiv \dfrac{\sigma_{\Delta T}}{\mu_{\Delta T}},
\end{align}
where $\sigma_{\Delta T}$ and $\mu_{\Delta T}$ are the standard deviation and mean of the time series, respectively, measured over some interval $\Delta T$. We use $\Delta T = 553\, GM/c^3\ (\approx 6.5$ months for M87*) to be consistent with the timescale used in the EHT analysis of the Galactic Center, which found inconsistencies between data and observation. MAD models are mostly unaffected since the geometric extent of the emission region does not change significantly with or without the helicity barrier. In contrast, in SANE models, especially with larger values of $\rhigh$, the emission tends to shift towards the disk when the helicity barrier is included, increasing the relative imprint of the turbulent dynamics near the horizon (which is more variable than in the funnel wall).

\section{Discussion}
\label{sec:discussion}

Our approach is subject to several limitations. First, our GRMHD simulations do not dynamically evolve electron temperatures and instead only track the total energy of the ion-plus-electron fluid, leaving the electron distribution function to be prescribed in post-processing. This procedure relies on the assumption that the ratio of heating rates can be directly mapped to the ratio of temperatures, which may be a reasonable assumption if the majority of internal energy is locally generated but need not be the case. Second, our base heating model does not depend on the structure of the accretion flow and does not correspond to any specific dissipation mechanism. Additionally, our simulations do not take into account the effects of pressure anisotropy on dynamics of the flow and thermodynamics of ions and electrons.

Additional limitations are due to our implementation of the helicity barrier physics. Although it is necessary to compute the strength of local fluctuations in the Els\"asser variables, there is no clear way to calculate the mean flow $\langle z_\pm \rangle$ due to the global geometric structure and the relativistic nature of the problem. In this work, we use temporal averages rather than spatial ones, and our results depend on the details of the averaging procedure. To estimate the uncertainty due to averaging, in Figure~\ref{fig:beta2_averaging_examples} we compare the measured value of $\beta_2$ for different averaging windows for a representative set of models and find that the choice is relatively robust. Our model assumes that inhomogeneities do not allow the locally generated helicity to be transported away.

Finally, even though the barrier is expected to form in low-$\beta$ regions of turbulence, we have applied the barrier-induced heating reductions across the entire domain. We do not expect this distinction to be qualitatively important, as high-$\beta$ regions have relatively small amounts of electron heating in any case. Nevertheless, the heating reduction due to helicity barrier applies only in the regions of the flow where the main dissipation mechanism is turbulence. In our simulations, we assume that the entire domain is dissipated through turbulence, which is most likely incorrect. The relative importance of different dissipation channels is not yet well understood.

\vspace{1em}

It is worthwhile to consider whether the effects of the helicity could be reproduced with modifications to the canonical electron temperature prescription of Equation~\ref{eqn:rhigh}. To first order, the helicity barrier produces cooler electrons across the domain and thereby increases the temperature ratio $T_i/T_e$ everywhere; this change could be emulated by increasing the $\rlow$ parameter by a factor of a few to order ten. Cool electron populations like the one that would result from this change have been invoked to explain disagreements between observations and model predictions for jet power and polarimetric properties \citep{EHTC_2019_5,EHTC_2021_8}.

Is it possible to do better? The strength of the helicity barrier depends on the normalized cross helicity $\sigma_c$. Comparing the panels of Figures~\ref{fig:sigmac_models} and \ref{fig:emission_source_comparison} shows that there is no clear relationship between $\sigmac$ and other fluid parameters, like $\sigma$ or $\beta$. Any modification to Equation~\ref{eqn:rhigh} would at least need to be a function of some other locally calculable quantity that is not readily identifiable, so it is not clear how to modify the prescription without introducing an extra complexity comparable to directly evaluating $\sigmac$.

Thus, while such a global approach would produce the same qualitative effects as incorporating the helicity barrier, the complicated structures seen in Figure~\ref{fig:average_vs_snapshot} suggest that any global approach would be inaccurate in detail. How the inaccuracies due to this approximation would compare to other modeling uncertainties is a different question, and we caution that the sensitive dependence of the observables on the details of electron distribution function makes it challenging to evaluate any kind of Jacobian. Performing a rigorous comparison is thus well beyond the scope of this paper.

\section{Summary}
\label{sec:summary}

We have studied the effects of imbalanced turbulence and the resultant helicity barrier in the context of radiative inefficient black hole accretion. We have computed the degree of cross-helicity buildup in a suite of numerical accretion simulations covering both magnetically arrested disk (MAD) and standard and normal evolution (SANE) flows and over a range of black hole spins. We have also used results from local simulations of non-relativistic low-$\beta$ turbulence \citep{Meyrand2021,Squire2022} to explore how including (or not) the helicity barrier in the imaging procedure can affect predictions for $230\,{\rm GHz}$ horizon-scale black hole images relevant for Event Horizon Telescope analyses \citep{EHTC_2019_5,EHTC_2021_8,EHTC_2022_5}. 

The local level of sustained imbalance determines the importance of the helicity barrier, which in turn limits electron heating. We have found that the imbalance tends to be smaller in regions of the flow with high plasma $\beta$ (commonly found in the disks of SANE flows and flows with low black hole spin). In contrast, in regions with ordered magnetic fields, such as in the jet and its surrounding wind in SANE flows as well as throughout much more of the domain in MAD flows, imbalance persists, helicity builds, and electron heating is more restricted. Accounting for the helicity barrier thus causes emission to shift away from the funnel wall towards the lower-magnetization disk region in SANE flows, while the emission morphology is largely unaffected in MADs. 

When comparing to observations, the total emission produced by a candidate accretion flow must match its observed value, and the cooler electrons require larger plasma number densities and magnetic field strengths. Thus, neglecting the helicity barrier can lead to underestimated accretion rates and inferred jet powers by more than a factor of two. The higher plasma densities also lead to increased Faraday depths and depolarization, resulting in decreased amplitudes of the polarimetric $\beta_2$ observable. Finally, we find that the inferred ring diameter and light curve variability modulation index are mostly unchanged for MAD flows but may increase for SANE flows, especially with large values of $\rhigh$. The increased jet powers and decreased coherent polarizations due to inclusion of the helicity barrier may help explain some qualitative differences between observed EHT data and contemporary modeling efforts \citep{EHTC_2019_5,EHTC_2021_8}.

\begin{acknowledgments}
The authors thank Michi Baub\"{o}ck, Andrew Chael, Matt Kunz, Elias Most, Eliot Quataert, Jonathan Squire, Jim Stone, and Muni Zhou for useful discussions and suggestions. The authors also thank the anonymous referee for useful comments and suggestions. G.N.W.~was supported by the Taplin Fellowship. Support for L.A.~was provided by the Institute for Advanced Study.
\end{acknowledgments}

\bibliography{main.bib,leva.bib}

\begin{thebibliography}{}
\expandafter\ifx\csname natexlab\endcsname\relax\def\natexlab#1{#1}\fi
\providecommand{\url}[1]{\href{#1}{#1}}
\providecommand{\dodoi}[1]{doi:~\href{http://doi.org/#1}{\nolinkurl{#1}}}
\providecommand{\doeprint}[1]{\href{http://ascl.net/#1}{\nolinkurl{http://ascl.net/#1}}}
\providecommand{\doarXiv}[1]{\href{https://arxiv.org/abs/#1}{\nolinkurl{https://arxiv.org/abs/#1}}}

\bibitem[{Akiyama {et~al.}(2015)Akiyama, Lu, Fish, Doeleman, Broderick, Dexter,
  Hada, Kino, Nagai, Honma, Johnson, Algaba, Asada, Brinkerink, Blundell,
  Bower, Cappallo, Crew, Dexter, Dzib, Freund, Friberg, Gurwell, Ho, Inoue,
  Krichbaum, Loinard, MacMahon, Marrone, Moran, Nakamura, Nagar, {Ortiz-Leon},
  Plambeck, Pradel, Primiani, Rogers, Roy, SooHoo, Tavares, Tilanus, Titus,
  Wagner, Weintroub, Yamaguchi, Young, Zensus, \&
  Ziurys}]{akiyama_2015_230GHzVLBI}
Akiyama, K., Lu, R.-S., Fish, V.~L., {et~al.} 2015, The Astrophysical Journal,
  807, 150, \dodoi{10.1088/0004-637X/807/2/150}

\bibitem[{Akiyama {et~al.}(2017)Akiyama, Kuramochi, Ikeda, Fish, Tazaki, Honma,
  Doeleman, E.~Broderick, Dexter, Mo{\'s}cibrodzka, Bouman, A.~Chael, \&
  Zaizen}]{akiyama_2017_ImagingSchwarzschildradiusscaleStructure}
Akiyama, K., Kuramochi, K., Ikeda, S., {et~al.} 2017, ApJ, 838, 1,
  \dodoi{10.3847/1538-4357/aa6305}

\bibitem[{{Arzamasskiy} {et~al.}(2019){Arzamasskiy}, {Kunz}, {Chandran}, \&
  {Quataert}}]{Arzamasskiy2019}
{Arzamasskiy}, L., {Kunz}, M.~W., {Chandran}, B.~D.~G., \& {Quataert}, E. 2019,
  \apj, 879, 53

\bibitem[{{Arzamasskiy} {et~al.}(2022){Arzamasskiy}, {Kunz}, {Squire},
  {Quataert}, \& {Schekochihin}}]{Arzamasskiy2022}
{Arzamasskiy}, L., {Kunz}, M.~W., {Squire}, J., {Quataert}, E., \&
  {Schekochihin}, A.~A. 2022, arXiv:2207.05189

\bibitem[{{Bacchini} {et~al.}(2022){Bacchini}, {Arzamasskiy}, {Zhdankin},
  {Werner}, {Begelman}, \& {Uzdensky}}]{Bacchini2022}
{Bacchini}, F., {Arzamasskiy}, L., {Zhdankin}, V., {et~al.} 2022, \apj, 938, 86

\bibitem[{Biskamp(2003)}]{biskamp_2003_mhdturbulence}
Biskamp, D. 2003, Magnetohydrodynamic {{Turbulence}} (Cambridge University
  Press)

\bibitem[{{Bisnovatyi-Kogan} \&
  Lovelace(1997)}]{bisnovatyikogan_1997_OhmicADAF}
{Bisnovatyi-Kogan}, G.~S., \& Lovelace, R. V.~E. 1997, The Astrophysical
  Journal, 486, L43, \dodoi{10.1086/310826}

\bibitem[{{Bisnovatyi-Kogan} \&
  Ruzmaikin(1974)}]{bisnovatyi-kogan_1974_AccretionMatterMAD}
{Bisnovatyi-Kogan}, G.~S., \& Ruzmaikin, A.~A. 1974, Astrophysics and Space
  Science, 28, 45, \dodoi{10.1007/BF00642237}

\bibitem[{Blandford \& Eichler(1987)}]{blandford_1987_shockacceleration}
Blandford, R., \& Eichler, D. 1987, Physics Reports, 154, 1,
  \dodoi{10.1016/0370-1573(87)90134-7}

\bibitem[{Boldyrev(2006)}]{boldyrev_2006_spectturb}
Boldyrev, S. 2006, Physical Review Letters, 96, 115002,
  \dodoi{10.1103/PhysRevLett.96.115002}

\bibitem[{{Bott} {et~al.}(2021){Bott}, {Arzamasskiy}, {Kunz}, {Quataert}, \&
  {Squire}}]{Bott2021}
{Bott}, A.~F.~A., {Arzamasskiy}, L., {Kunz}, M.~W., {Quataert}, E., \&
  {Squire}, J. 2021, \apjl, 922, L35

\bibitem[{{Bott} {et~al.}(2023){Bott}, {Kunz}, {Quataert}, {Arzamasskiy}, \&
  {Squire}}]{Bott2023}
{Bott}, A.~F.~A., {Kunz}, M.~W., {Quataert}, E., {Arzamasskiy}, L., \&
  {Squire}, J. 2023, in preparation

\bibitem[{{Bowen} {et~al.}(2023){Bowen}, {Bale}, {Chandran}, {Chasapis},
  {Chen}, {Dudok de Wit}, {Mallet}, {Meyrand}, \&
  {Squire}}]{bowen_2023_MediationCollisionlessTurbulentDissipation}
{Bowen}, T.~A., {Bale}, S.~D., {Chandran}, B. D.~G., {et~al.} 2023, arXiv
  e-prints, arXiv:2306.04881, \dodoi{10.48550/arXiv.2306.04881}

\bibitem[{{Cerri} {et~al.}(2021){Cerri}, {Arzamasskiy}, \& {Kunz}}]{Cerri2021}
{Cerri}, S.~S., {Arzamasskiy}, L., \& {Kunz}, M.~W. 2021, \apj, 916, 120

\bibitem[{Chael {et~al.}(2018)Chael, Rowan, Narayan, Johnson, \&
  Sironi}]{chael_2018_RoleElectronHeating}
Chael, A., Rowan, M., Narayan, R., Johnson, M., \& Sironi, L. 2018, Monthly
  Notices of the Royal Astronomical Society, 478, 5209,
  \dodoi{10.1093/mnras/sty1261}

\bibitem[{Chael {et~al.}(2022)Chael, Chan, {Klbouman}, Wielgus, Farah, Palumbo,
  Blackburn, {Aviad}, {Dpesce}, Quarles, Akiyama, {Michaeldjohnson}, Wong, Bao,
  Tiede, Aldinger, Clauss, Grayson, Junior, Romero, Lindahl, {Ilya}, Cheong,
  Rawat, \& {ZFudge}}]{chael_2022_ehtim}
Chael, A., Chan, C.-K., {Klbouman}, {et~al.} 2022, Zenodo,
  \dodoi{10.5281/zenodo.6519440}

\bibitem[{Chandran {et~al.}(2018)Chandran, Foucart, \&
  Tchekhovskoy}]{chandran_2018_heatgrmhdturb}
Chandran, B. D.~G., Foucart, F., \& Tchekhovskoy, A. 2018, Journal of Plasma
  Physics, 84, 905840310, \dodoi{10.1017/S0022377818000387}

\bibitem[{{Chandran} {et~al.}(2010){Chandran}, {Li}, {Rogers}, {Quataert}, \&
  {Germaschewski}}]{Chandran2010}
{Chandran}, B.~D.~G., {Li}, B., {Rogers}, B.~N., {Quataert}, E., \&
  {Germaschewski}, K. 2010, \apj, 720, 503

\bibitem[{Chandrasekhar(1960)}]{chandrasekhar_1960_RadiativeTransfer}
Chandrasekhar, S. 1960, Radiative Transfer ({Dover})

\bibitem[{{Comisso} \& {Sironi}(2018)}]{Comisso2018}
{Comisso}, L., \& {Sironi}, L. 2018, \prl, 121, 255101

\bibitem[{{Comisso} \& {Sironi}(2019)}]{Comisso2019}
---. 2019, \apj, 886, 122

\bibitem[{De~Villiers \& Hawley(2003)}]{devilliers_2003_grmhdtori}
De~Villiers, J.-P., \& Hawley, J.~F. 2003, The Astrophysical Journal, 592,
  1060, \dodoi{10.1086/375866}

\bibitem[{Doeleman {et~al.}(2012)Doeleman, Fish, Schenck, Beaudoin, Blundell,
  Bower, Broderick, Chamberlin, Freund, Friberg, Gurwell, Ho, Honma, Inoue,
  Krichbaum, Lamb, Loeb, Lonsdale, Marrone, Moran, Oyama, Plambeck, Primiani,
  Rogers, Smythe, SooHoo, Strittmatter, Tilanus, Titus, Weintroub, Wright,
  Young, \& Ziurys}]{doeleman_2012_JetLaunchingStructureResolved}
Doeleman, S.~S., Fish, V.~L., Schenck, D.~E., {et~al.} 2012, Science, 338, 355,
  \dodoi{10.1126/science.1224768}

\bibitem[{{Event Horizon Telescope Collaboration}
  {et~al.}(2019{\natexlab{a}}){Event Horizon Telescope Collaboration}, Akiyama,
  Alberdi, Alef, Asada, Azulay, Baczko, Ball, Balokovi{\'c}, Barrett, Bintley,
  Blackburn, Boland, Bouman, Bower, Bremer, Brinkerink, Brissenden, Britzen,
  Broderick, Broguiere, Bronzwaer, Byun, Carlstrom, Chael, Chan, Chatterjee,
  Chatterjee, Chen, Chen, Cho, Christian, Conway, Cordes, Crew, Cui, Davelaar,
  Laurentis, Deane, Dempsey, Desvignes, Dexter, Doeleman, Eatough, Falcke,
  Fish, Fomalont, {Fraga-Encinas}, Freeman, Friberg, Fromm, G{\'o}mez, Galison,
  Gammie, Garc{\'i}a, Gentaz, Georgiev, Goddi, Gold, Gu, Gurwell, Hada, Hecht,
  Hesper, Ho, Ho, Honma, Huang, Huang, Hughes, Ikeda, Inoue, Issaoun, James,
  Jannuzi, Janssen, Jeter, Jiang, Johnson, Jorstad, Jung, Karami, Karuppusamy,
  Kawashima, Keating, Kettenis, Kim, Kim, Kim, Kino, Koay, Koch, Koyama,
  Kramer, Kramer, Krichbaum, Kuo, Lauer, Lee, Li, Li, Lindqvist, Liu, Liuzzo,
  Lo, Lobanov, Loinard, Lonsdale, Lu, MacDonald, Mao, Markoff, Marrone,
  Marscher, {Mart{\'i}-Vidal}, Matsushita, Matthews, Medeiros, Menten, Mizuno,
  Mizuno, Moran, Moriyama, Moscibrodzka, M{\"u}ller, Nagai, Nagar, Nakamura,
  Narayan, Narayanan, Natarajan, Neri, Ni, Noutsos, Okino, Olivares,
  {Ortiz-Le{\'o}n}, Oyama, {\"O}zel, Palumbo, Patel, Pen, Pesce, Pi{\'e}tu,
  Plambeck, PopStefanija, Porth, Prather, {Preciado-L{\'o}pez}, Psaltis, Pu,
  Ramakrishnan, Rao, Rawlings, Raymond, Rezzolla, Ripperda, Roelofs, Rogers,
  Ros, Rose, Roshanineshat, Rottmann, Roy, Ruszczyk, Ryan, Rygl, S{\'a}nchez,
  {S{\'a}nchez-Arguelles}, Sasada, Savolainen, Schloerb, Schuster, Shao, Shen,
  Small, Sohn, SooHoo, Tazaki, Tiede, Tilanus, Titus, Toma, Torne, Trent,
  Trippe, Tsuda, van Bemmel, van Langevelde, van Rossum, Wagner, Wardle,
  Weintroub, Wex, Wharton, Wielgus, Wong, Wu, Young, Young, Younsi, Yuan, Yuan,
  Zensus, Zhao, Zhao, Zhu, Algaba, Allardi, Amestica, Anczarski, Bach,
  Baganoff, Beaudoin, Benson, Berthold, Blanchard, Blundell, Bustamente,
  Cappallo, {Castillo-Dom{\'i}nguez}, Chang, Chang, Chang, Chen, Chilson,
  Chuter, Rosado, Coulson, Crawford, Crowley, David, Derome, Dexter, Dornbusch,
  Dudevoir, Dzib, Eckart, Eckert, Erickson, Everett, Faber, Farah, Fath,
  Folkers, Forbes, Freund, {G{\'o}mez-Ruiz}, Gale, Gao, Geertsema, Graham,
  Greer, Grosslein, Gueth, Haggard, Halverson, Han, Han, Hao, Hasegawa,
  Henning, {Hern{\'a}ndez-G{\'o}mez}, {Herrero-Illana}, Heyminck, Hirota, Hoge,
  Huang, Impellizzeri, Jiang, Kamble, Keisler, Kimura, Kono, Kubo, Kuroda,
  Lacasse, Laing, Leitch, Li, Lin, Liu, Liu, Lu, Marson, {Martin-Cocher},
  Massingill, Matulonis, McColl, McWhirter, Messias, {Meyer-Zhao}, Michalik,
  Monta{\~n}a, Montgomerie, {Mora-Klein}, Muders, Nadolski, Navarro, Neilsen,
  Nguyen, Nishioka, Norton, Nowak, Nystrom, Ogawa, Oshiro, Oyama, Parsons,
  Paine, Pe{\~n}alver, Phillips, Poirier, Pradel, Primiani, Raffin, Rahlin,
  Reiland, Risacher, Ruiz, {S{\'a}ez-Mada{\'i}n}, Sassella, Schellart, Shaw,
  Silva, Shiokawa, Smith, Snow, Souccar, Sousa, Sridharan, Srinivasan, Stahm,
  Stark, Story, Timmer, Vertatschitsch, Walther, Wei, Whitehorn, Whitney,
  Woody, Wouterloot, Wright, Yamaguchi, Yu, Zeballos, Zhang, \&
  Ziurys}]{EHTC_2019_1}
{Event Horizon Telescope Collaboration}, Akiyama, K., Alberdi, A., {et~al.}
  2019{\natexlab{a}}, ApJL, 875, L1, \dodoi{10.3847/2041-8213/ab0ec7}

\bibitem[{{Event Horizon Telescope Collaboration}
  {et~al.}(2019{\natexlab{b}}){Event Horizon Telescope Collaboration}, Akiyama,
  Alberdi, Alef, Asada, Azulay, Baczko, Ball, Balokovi{\'c}, Barrett, Bintley,
  Blackburn, Boland, Bouman, Bower, Bremer, Brinkerink, Brissenden, Britzen,
  Broderick, Broguiere, Bronzwaer, Byun, Carlstrom, Chael, Chan, Chatterjee,
  Chatterjee, Chen, Chen, Cho, Christian, Conway, Cordes, Crew, Cui, Davelaar,
  Laurentis, Deane, Dempsey, Desvignes, Dexter, Doeleman, Eatough, Falcke,
  Fish, Fomalont, {Fraga-Encinas}, Friberg, Fromm, G{\'o}mez, Galison, Gammie,
  Garc{\'i}a, Gentaz, Georgiev, Goddi, Gold, Gu, Gurwell, Hada, Hecht, Hesper,
  Ho, Ho, Honma, Huang, Huang, Hughes, Ikeda, Inoue, Issaoun, James, Jannuzi,
  Janssen, Jeter, Jiang, Johnson, Jorstad, Jung, Karami, Karuppusamy,
  Kawashima, Keating, Kettenis, Kim, Kim, Kim, Kino, Koay, Koch, Koyama,
  Kramer, Kramer, Krichbaum, Kuo, Lauer, Lee, Li, Li, Lindqvist, Liu, Liuzzo,
  Lo, Lobanov, Loinard, Lonsdale, Lu, MacDonald, Mao, Markoff, Marrone,
  Marscher, {Mart{\'i}-Vidal}, Matsushita, Matthews, Medeiros, Menten, Mizuno,
  Mizuno, Moran, Moriyama, Moscibrodzka, Mu{\"l}ler, Nagai, Nagar, Nakamura,
  Narayan, Narayanan, Natarajan, Neri, Ni, Noutsos, Okino, Olivares, Oyama,
  {\"O}zel, Palumbo, Patel, Pen, Pesce, Pi{\'e}tu, Plambeck, PopStefanija,
  Porth, Prather, {Preciado-L{\'o}pez}, Psaltis, Pu, Ramakrishnan, Rao,
  Rawlings, Raymond, Rezzolla, Ripperda, Roelofs, Rogers, Ros, Rose,
  Roshanineshat, Rottmann, Roy, Ruszczyk, Ryan, Rygl, S{\'a}nchez,
  {S{\'a}nchez-Arguelles}, Sasada, Savolainen, Schloerb, Schuster, Shao, Shen,
  Small, Sohn, SooHoo, Tazaki, Tiede, Tilanus, Titus, Toma, Torne, Trent,
  Trippe, Tsuda, van Bemmel, van Langevelde, van Rossum, Wagner, Wardle,
  Weintroub, Wex, Wharton, Wielgus, Wong, Wu, Young, Young, Younsi, Yuan, Yuan,
  Zensus, Zhao, Zhao, Zhu, Anczarski, Baganoff, Eckart, Farah, Haggard,
  {Meyer-Zhao}, Michalik, Nadolski, Neilsen, Nishioka, Nowak, Pradel, Primiani,
  Souccar, Vertatschitsch, Yamaguchi, \& Zhang}]{EHTC_2019_5}
---. 2019{\natexlab{b}}, ApJL, 875, L5, \dodoi{10.3847/2041-8213/ab0f43}

\bibitem[{{Event Horizon Telescope Collaboration}
  {et~al.}(2019{\natexlab{c}}){Event Horizon Telescope Collaboration}, Akiyama,
  Alberdi, Alef, Asada, Azulay, Baczko, Ball, Balokovi{\'c}, Barrett, Bintley,
  Blackburn, Boland, Bouman, Bower, Bremer, Brinkerink, Brissenden, Britzen,
  Broderick, Broguiere, Bronzwaer, Byun, Carlstrom, Chael, Chan, Chatterjee,
  Chatterjee, Chen, Chen, Cho, Christian, Conway, Cordes, Crew, Cui, Davelaar,
  Laurentis, Deane, Dempsey, Desvignes, Dexter, Doeleman, Eatough, Falcke,
  Fish, Fomalont, {Fraga-Encinas}, Friberg, Fromm, G{\'o}mez, Galison, Gammie,
  Garc{\'i}a, Gentaz, Georgiev, Goddi, Gold, Gu, Gurwell, Hada, Hecht, Hesper,
  Ho, Ho, Honma, Huang, Huang, Hughes, Ikeda, Inoue, Issaoun, James, Jannuzi,
  Janssen, Jeter, Jiang, Johnson, Jorstad, Jung, Karami, Karuppusamy,
  Kawashima, Keating, Kettenis, Kim, Kim, Kim, Kino, Koay, Koch, Koyama,
  Kramer, Kramer, Krichbaum, Kuo, Lauer, Lee, Li, Li, Lindqvist, Liu, Liuzzo,
  Lo, Lobanov, Loinard, Lonsdale, Lu, MacDonald, Mao, Markoff, Marrone,
  Marscher, {Mart{\'i}-Vidal}, Matsushita, Matthews, Medeiros, Menten, Mizuno,
  Mizuno, Moran, Moriyama, Moscibrodzka, M{\"u}ller, Nagai, Nagar, Nakamura,
  Narayan, Narayanan, Natarajan, Neri, Ni, Noutsos, Okino, Olivares, Oyama,
  {\"O}zel, Palumbo, Patel, Pen, Pesce, Pi{\'e}tu, Plambeck, PopStefanija,
  Porth, Prather, {Preciado-L{\'o}pez}, Psaltis, Pu, Ramakrishnan, Rao,
  Rawlings, Raymond, Rezzolla, Ripperda, Roelofs, Rogers, Ros, Rose,
  Roshanineshat, Rottmann, Roy, Ruszczyk, Ryan, Rygl, S{\'a}nchez,
  {S{\'a}nchez-Arguelles}, Sasada, Savolainen, Schloerb, Schuster, Shao, Shen,
  Small, Sohn, SooHoo, Tazaki, Tiede, Tilanus, Titus, Toma, Torne, Trent,
  Trippe, Tsuda, van Bemmel, van Langevelde, van Rossum, Wagner, Wardle,
  Weintroub, Wex, Wharton, Wielgus, Wong, Wu, Young, Young, Younsi, Yuan, Yuan,
  Zensus, Zhao, Zhao, Zhu, Farah, {Meyer-Zhao}, Michalik, Nadolski, Nishioka,
  Pradel, Primiani, Souccar, Vertatschitsch, \& Yamaguchi}]{EHTC_2019_6}
---. 2019{\natexlab{c}}, ApJL, 875, L6, \dodoi{10.3847/2041-8213/ab1141}

\bibitem[{{Event Horizon Telescope Collaboration}
  {et~al.}(2019{\natexlab{d}}){Event Horizon Telescope Collaboration}, Akiyama,
  Alberdi, Alef, Asada, Azulay, Baczko, Ball, Balokovi{\'c}, Barrett, Bintley,
  Blackburn, Boland, Bouman, Bower, Bremer, Brinkerink, Brissenden, Britzen,
  Broderick, Broguiere, Bronzwaer, Byun, Carlstrom, Chael, Chan, Chatterjee,
  Chatterjee, Chen, Chen, Cho, Christian, Conway, Cordes, Crew, Cui, Davelaar,
  Laurentis, Deane, Dempsey, Desvignes, Dexter, Doeleman, Eatough, Falcke,
  Fish, Fomalont, {Fraga-Encinas}, Freeman, Friberg, Fromm, G{\'o}mez, Galison,
  Gammie, Garc{\'i}a, Gentaz, Georgiev, Goddi, Gold, Gu, Gurwell, Hada, Hecht,
  Hesper, Ho, Ho, Honma, Huang, Huang, Hughes, Ikeda, Inoue, Issaoun, James,
  Jannuzi, Janssen, Jeter, Jiang, Johnson, Jorstad, Jung, Karami, Karuppusamy,
  Kawashima, Keating, Kettenis, Kim, Kim, Kim, Kino, Koay, Koch, Koyama,
  Kramer, Kramer, Krichbaum, Kuo, Lauer, Lee, Li, Li, Lindqvist, Liu, Liuzzo,
  Lo, Lobanov, Loinard, Lonsdale, Lu, MacDonald, Mao, Markoff, Marrone,
  Marscher, {Mart{\'i}-Vidal}, Matsushita, Matthews, Medeiros, Menten, Mizuno,
  Mizuno, Moran, Moriyama, Moscibrodzka, M{\"u}ller, Nagai, Nagar, Nakamura,
  Narayan, Narayanan, Natarajan, Neri, Ni, Noutsos, Okino, Olivares, Oyama,
  {\"O}zel, Palumbo, Patel, Pen, Pesce, Pi{\'e}tu, Plambeck, PopStefanija,
  Porth, Prather, {Preciado-L{\'o}pez}, Psaltis, Pu, Ramakrishnan, Rao,
  Rawlings, Raymond, Rezzolla, Ripperda, Roelofs, Rogers, Ros, Rose,
  Roshanineshat, Rottmann, Roy, Ruszczyk, Ryan, Rygl, S{\'a}nchez,
  {S{\'a}nchez-Arguelles}, Sasada, Savolainen, Schloerb, Schuster, Shao, Shen,
  Small, Sohn, SooHoo, Tazaki, Tiede, Tilanus, Titus, Toma, Torne, Trent,
  Trippe, Tsuda, van Bemmel, van Langevelde, van Rossum, Wagner, Wardle,
  Weintroub, Wex, Wharton, Wielgus, Wong, Wu, Young, Young, Younsi, Yuan, Yuan,
  Zensus, Zhao, Zhao, Zhu, Farah, {Meyer-Zhao}, Michalik, Nadolski, Nishioka,
  Pradel, Primiani, Souccar, Vertatschitsch, \& Yamaguchi}]{EHTC_2019_4}
---. 2019{\natexlab{d}}, ApJL, 875, L4, \dodoi{10.3847/2041-8213/ab0e85}

\bibitem[{{Event Horizon Telescope Collaboration}
  {et~al.}(2021{\natexlab{a}}){Event Horizon Telescope Collaboration}, Akiyama,
  Algaba, Alberdi, Alef, Anantua, Asada, Azulay, Baczko, Ball, Balokovi{\'c},
  Barrett, Benson, Bintley, Blackburn, Blundell, Boland, Bouman, Bower, Boyce,
  Bremer, Brinkerink, Brissenden, Britzen, Broderick, Broguiere, Bronzwaer,
  Byun, Carlstrom, Chael, Chan, Chatterjee, Chatterjee, Chen, Chen, Chesler,
  Cho, Christian, Conway, Cordes, Crawford, Crew, {Cruz-Osorio}, Cui, Davelaar,
  Laurentis, Deane, Dempsey, Desvignes, Dexter, Doeleman, Eatough, Falcke,
  Farah, Fish, Fomalont, Ford, {Fraga-Encinas}, Freeman, Friberg, Fromm,
  Fuentes, Galison, Gammie, Garc{\'i}a, Gentaz, Georgiev, Goddi, Gold,
  G{\'o}mez, {G{\'o}mez-Ruiz}, Gu, Gurwell, Hada, Haggard, Hecht, Hesper, Ho,
  Ho, Honma, Huang, Huang, Hughes, Ikeda, Inoue, Issaoun, James, Jannuzi,
  Janssen, Jeter, Jiang, {Jimenez-Rosales}, Johnson, Jorstad, Jung, Karami,
  Karuppusamy, Kawashima, Keating, Kettenis, Kim, Kim, Kim, Kim, Kino, Koay,
  Kofuji, Koch, Koyama, Kramer, Kramer, Krichbaum, Kuo, Lauer, Lee, Levis, Li,
  Li, Lindqvist, Lico, Lindahl, Liu, Liu, Liuzzo, Lo, Lobanov, Loinard,
  Lonsdale, Lu, MacDonald, Mao, Marchili, Markoff, Marrone, Marscher,
  {Mart{\'i}-Vidal}, Matsushita, Matthews, Medeiros, Menten, Mizuno, Mizuno,
  Moran, Moriyama, Moscibrodzka, M{\"u}ller, Musoke, Mej{\'i}as, Michalik,
  Nadolski, Nagai, Nagar, Nakamura, Narayan, Narayanan, Natarajan, Nathanail,
  Neilsen, Neri, Ni, Noutsos, Nowak, Okino, Olivares, {Ortiz-Le{\'o}n}, Oyama,
  {\"O}zel, Palumbo, Park, Patel, Pen, Pesce, Pi{\'e}tu, Plambeck,
  PopStefanija, Porth, P{\"o}tzl, Prather, {Preciado-L{\'o}pez}, Psaltis, Pu,
  Ramakrishnan, Rao, Rawlings, Raymond, Rezzolla, Ricarte, Ripperda, Roelofs,
  Rogers, Ros, Rose, Roshanineshat, Rottmann, Roy, Ruszczyk, Rygl, S{\'a}nchez,
  {S{\'a}nchez-Arguelles}, Sasada, Savolainen, Schloerb, Schuster, Shao, Shen,
  Small, Sohn, SooHoo, Sun, Tazaki, Tetarenko, Tiede, Tilanus, Titus, Toma,
  Torne, Trent, Traianou, Trippe, van Bemmel, van Langevelde, van Rossum,
  Wagner, {Ward-Thompson}, Wardle, Weintroub, Wex, Wharton, Wielgus, Wong, Wu,
  Yoon, Young, Young, Younsi, Yuan, Yuan, Zensus, Zhao, Zhao, \&
  Collaboration}]{EHTC_2021_7}
{Event Horizon Telescope Collaboration}, Akiyama, K., Algaba, J.~C., {et~al.}
  2021{\natexlab{a}}, ApJL, 910, L12, \dodoi{10.3847/2041-8213/abe71d}

\bibitem[{{Event Horizon Telescope Collaboration}
  {et~al.}(2021{\natexlab{b}}){Event Horizon Telescope Collaboration}, Akiyama,
  Algaba, Alberdi, Alef, Anantua, Asada, Azulay, Baczko, Ball, Balokovi{\'c},
  Barrett, Benson, Bintley, Blackburn, Blundell, Boland, Bouman, Bower, Boyce,
  Bremer, Brinkerink, Brissenden, Britzen, Broderick, Broguiere, Bronzwaer,
  Byun, Carlstrom, Chael, Chan, Chatterjee, Chatterjee, Chen, Chen, Chesler,
  Cho, Christian, Conway, Cordes, Crawford, Crew, {Cruz-Osorio}, Cui, Davelaar,
  Laurentis, Deane, Dempsey, Desvignes, Dexter, Doeleman, Eatough, Falcke,
  Farah, Fish, Fomalont, Ford, {Fraga-Encinas}, Friberg, Fromm, Fuentes,
  Galison, Gammie, Garc{\'i}a, Gelles, Gentaz, Georgiev, Goddi, Gold,
  G{\'o}mez, {G{\'o}mez-Ruiz}, Gu, Gurwell, Hada, Haggard, Hecht, Hesper,
  Himwich, Ho, Ho, Honma, Huang, Huang, Hughes, Ikeda, Inoue, Issaoun, James,
  Jannuzi, Janssen, Jeter, Jiang, {Jimenez-Rosales}, Johnson, Jorstad, Jung,
  Karami, Karuppusamy, Kawashima, Keating, Kettenis, Kim, Kim, Kim, Kim, Kino,
  Koay, Kofuji, Koch, Koyama, Kramer, Kramer, Krichbaum, Kuo, Lauer, Lee,
  Levis, Li, Li, Lindqvist, Lico, Lindahl, Liu, Liu, Liuzzo, Lo, Lobanov,
  Loinard, Lonsdale, Lu, MacDonald, Mao, Marchili, Markoff, Marrone, Marscher,
  {Mart{\'i}-Vidal}, Matsushita, Matthews, Medeiros, Menten, Mizuno, Mizuno,
  Moran, Moriyama, Moscibrodzka, M{\"u}ller, Musoke, Mej{\'i}as, Michalik,
  Nadolski, Nagai, Nagar, Nakamura, Narayan, Narayanan, Natarajan, Nathanail,
  Neilsen, Neri, Ni, Noutsos, Nowak, Okino, Olivares, {Ortiz-Le{\'o}n}, Oyama,
  {\"O}zel, Palumbo, Park, Patel, Pen, Pesce, Pi{\'e}tu, Plambeck,
  PopStefanija, Porth, P{\"o}tzl, Prather, {Preciado-L{\'o}pez}, Psaltis, Pu,
  Ramakrishnan, Rao, Rawlings, Raymond, Rezzolla, Ricarte, Ripperda, Roelofs,
  Rogers, Ros, Rose, Roshanineshat, Rottmann, Roy, Ruszczyk, Rygl, S{\'a}nchez,
  {S{\'a}nchez-Arguelles}, Sasada, Savolainen, Schloerb, Schuster, Shao, Shen,
  Small, Sohn, SooHoo, Sun, Tazaki, Tetarenko, Tiede, Tilanus, Titus, Toma,
  Torne, Trent, Traianou, Trippe, van Bemmel, van Langevelde, van Rossum,
  Wagner, {Ward-Thompson}, Wardle, Weintroub, Wex, Wharton, Wielgus, Wong, Wu,
  Yoon, Young, Young, Younsi, Yuan, Yuan, Zensus, Zhao, Zhao, \&
  Collaboration}]{EHTC_2021_8}
---. 2021{\natexlab{b}}, ApJL, 910, L13, \dodoi{10.3847/2041-8213/abe4de}

\bibitem[{{Event Horizon Telescope Collaboration}
  {et~al.}(2022{\natexlab{a}}){Event Horizon Telescope Collaboration}, Akiyama,
  Alberdi, Alef, Algaba, Anantua, Asada, Azulay, Bach, Baczko, Ball,
  Balokovi{\'c}, Barrett, Baub{\"o}ck, Benson, Bintley, Blackburn, Blundell,
  Bouman, Bower, Boyce, Bremer, Brinkerink, Brissenden, Britzen, Broderick,
  Broguiere, Bronzwaer, Bustamante, Byun, Carlstrom, Ceccobello, Chael, Chan,
  Chatterjee, Chatterjee, Chen, Chen, Cheng, Cho, Christian, Conroy, Conway,
  Cordes, Crawford, Crew, {Cruz-Osorio}, Cui, Davelaar, De~Laurentis, Deane,
  Dempsey, Desvignes, Dexter, Dhruv, Doeleman, Dougal, Dzib, Eatough, Emami,
  Falcke, Farah, Fish, Fomalont, Ford, {Fraga-Encinas}, Freeman, Friberg,
  Fromm, Fuentes, Galison, Gammie, Garc{\'i}a, Gentaz, Georgiev, Goddi, Gold,
  {G{\'o}mez-Ruiz}, G{\'o}mez, Gu, Gurwell, Hada, Haggard, Haworth, Hecht,
  Hesper, Heumann, Ho, Ho, Honma, Huang, Huang, Hughes, Ikeda, Impellizzeri,
  Inoue, Issaoun, James, Jannuzi, Janssen, Jeter, Jiang, {Jim{\'e}nez-Rosales},
  Johnson, Jorstad, Joshi, Jung, Karami, Karuppusamy, Kawashima, Keating,
  Kettenis, Kim, Kim, Kim, Kim, Kino, Koay, Kocherlakota, Kofuji, Koch, Koyama,
  Kramer, Kramer, Krichbaum, Kuo, La~Bella, Lauer, Lee, Lee, Leung, Levis, Li,
  Lico, Lindahl, Lindqvist, Lisakov, Liu, Liu, Liuzzo, Lo, Lobanov, Loinard,
  Lonsdale, Lu, Mao, Marchili, Markoff, Marrone, Marscher, {Mart{\'i}-Vidal},
  Matsushita, Matthews, Medeiros, Menten, Michalik, Mizuno, Mizuno, Moran,
  Moriyama, Moscibrodzka, M{\"u}ller, Mus, Musoke, Myserlis, Nadolski, Nagai,
  Nagar, Nakamura, Narayan, Narayanan, Natarajan, Nathanail, Navarro~Fuentes,
  Neilsen, Neri, Ni, Noutsos, Nowak, Oh, Okino, Olivares, {Ortiz-Le{\'o}n},
  Oyama, {\"O}zel, Palumbo, Paraschos, Park, Parsons, Patel, Pen, Pesce,
  Pi{\'e}tu, Plambeck, PopStefanija, Porth, P{\"o}tzl, Prather,
  {Preciado-L{\'o}pez}, Psaltis, Pu, Ramakrishnan, Rao, Rawlings, Raymond,
  Rezzolla, Ricarte, Ripperda, Roelofs, Rogers, Ros, {Romero-Ca{\~n}izales},
  Roshanineshat, Rottmann, Roy, Ruiz, Ruszczyk, Rygl, S{\'a}nchez,
  {S{\'a}nchez-Arg{\"u}elles}, {S{\'a}nchez-Portal}, Sasada, Satapathy,
  Savolainen, Schloerb, Schonfeld, Schuster, Shao, Shen, Small, Sohn, SooHoo,
  Souccar, Sun, Tazaki, Tetarenko, Tiede, Tilanus, Titus, Torne, Traianou,
  Trent, Trippe, Turk, {van Bemmel}, {van Langevelde}, {van Rossum}, Vos,
  Wagner, {Ward-Thompson}, Wardle, Weintroub, Wex, Wharton, Wielgus, Wiik,
  Witzel, Wondrak, Wong, Wu, Yamaguchi, Yoon, Young, Young, Younsi, Yuan, Yuan,
  Zensus, Zhang, Zhao, Zhao, Agurto, Allardi, Amestica, Araneda, Arriagada,
  Berghuis, Bertarini, Berthold, Blanchard, Brown, C{\'a}rdenas, Cantzler,
  Caro, {Castillo-Dom{\'i}nguez}, Chan, Chang, Chang, Chang, Chang, Chen,
  Chilson, Chuter, Ciechanowicz, {Colin-Beltran}, Coulson, Crowley, Degenaar,
  Dornbusch, Dur{\'a}n, Everett, Faber, Forster, Fuchs, Gale, Geertsema,
  Gonz{\'a}lez, Graham, Gueth, Halverson, Han, Han, Hasegawa,
  {Hern{\'a}ndez-Rebollar}, Herrera, {Herrero-Illana}, Heyminck, Hirota, Hoge,
  Hostler~Schimpf, Howie, Huang, Jiang, Jinchi, John, Kimura, Klein, Kubo,
  Kuroda, Kwon, Lacasse, Laing, Leitch, Li, Liu, Liu, Lin, Lu, {Mac-Auliffe},
  {Martin-Cocher}, Matulonis, Maute, Messias, {Meyer-Zhao}, Monta{\~n}a,
  {Montenegro-Montes}, Montgomerie, Moreno~Nolasco, Muders, Nishioka, Norton,
  Nystrom, Ogawa, Olivares, Oshiro, {P{\'e}rez-Beaupuits}, Parra, Phillips,
  Poirier, Pradel, Qiu, Raffin, Rahlin, Ram{\'i}rez, Ressler, Reynolds,
  {Rodr{\'i}guez-Montoya}, {Saez-Madain}, Santana, Shaw, Shirkey, Silva, Snow,
  Sousa, Sridharan, Stahm, Stark, Test, Torstensson, Venegas, Walther, Wei,
  White, Wieching, Wijnands, Wouterloot, Yu, Yu, \& Zeballos}]{EHTC_2022_1}
{Event Horizon Telescope Collaboration}, Akiyama, K., Alberdi, A., {et~al.}
  2022{\natexlab{a}}, The Astrophysical Journal, 930, L12,
  \dodoi{10.3847/2041-8213/ac6674}

\bibitem[{{Event Horizon Telescope Collaboration}
  {et~al.}(2022{\natexlab{b}}){Event Horizon Telescope Collaboration}, Akiyama,
  Alberdi, Alef, Carlos~Algaba, Anantua, Asada, Azulay, Bach, Baczko, Ball,
  {Mislav Balokovi\'c}, Barrett, Baub{\"o}ck, Benson, Bintley, Blackburn,
  Blundell, Bouman, Bower, Boyce, Bremer, Brinkerink, Brissenden, Britzen,
  Broderick, Broguiere, Bronzwaer, Bustamante, Byun, Carlstrom, Ceccobello,
  Chael, Chan, Chatterjee, Chatterjee, Chen, Chen, Cheng, Cho, Christian,
  Conroy, Conway, Cordes, Crawford, Crew, {Cruz-Osorio}, Cui, Davelaar,
  De~Laurentis, Deane, Dempsey, Desvignes, Dexter, Dhruv, Doeleman, Dougal,
  Dzib, Eatough, Emami, Falcke, Farah, Fish, Fomalont, Ford, {Fraga-Encinas},
  Freeman, Friberg, Fromm, Fuentes, Galison, Gammie, Garc{\'i}a, Gentaz,
  Georgiev, Goddi, Gold, {G{\'o}mez-Ruiz}, G{\'o}mez, Gu, Gurwell, Hada,
  Haggard, Haworth, Hecht, Hesper, Heumann, Ho, Ho, Honma, Huang, Huang,
  Hughes, Ikeda, Violette~Impellizzeri, Inoue, Issaoun, James, Jannuzi,
  Janssen, Jeter, Jiang, {Jim{\'e}nez-Rosales}, Johnson, Jorstad, Joshi, Jung,
  Karami, Karuppusamy, Kawashima, Keating, Kettenis, Kim, Kim, Kim, Kim, Kino,
  Koay, Kocherlakota, Kofuji, Koch, Koyama, Kramer, Kramer, Krichbaum, Kuo,
  Bella, Lauer, Lee, Lee, Leung, Levis, Li, Lico, Lindahl, Lindqvist, Lisakov,
  Liu, Liu, Liuzzo, Lo, Lobanov, Loinard, Lonsdale, Lu, Mao, Marchili, Markoff,
  Marrone, Marscher, {Mart{\'i}-Vidal}, Matsushita, Matthews, Medeiros, Menten,
  Michalik, Mizuno, Mizuno, Moran, Moriyama, Moscibrodzka, M{\"u}ller, Mus,
  Musoke, Myserlis, Nadolski, Nagai, Nagar, Nakamura, Narayan, Narayanan,
  Natarajan, Nathanail, Navarro~Fuentes, Neilsen, Neri, Ni, Noutsos, Nowak, Oh,
  Okino, Olivares, {Ortiz-Le{\'o}n}, Oyama, {\"O}zel, Palumbo,
  Filippos~Paraschos, Park, Parsons, Patel, Pen, Pesce, Pi{\'e}tu, Plambeck,
  PopStefanija, Porth, P{\"o}tzl, Prather, {Preciado-L{\'o}pez}, Psaltis, Pu,
  Ramakrishnan, Rao, Rawlings, Raymond, Rezzolla, Ricarte, Ripperda, Roelofs,
  Rogers, Ros, {Romero-Ca{\~n}izales}, Roshanineshat, Rottmann, Roy, Ruiz,
  Ruszczyk, Rygl, S{\'a}nchez, {S{\'a}nchez-Arg{\"u}elles},
  {S{\'a}nchez-Portal}, Sasada, Satapathy, Savolainen, Schloerb, Schonfeld,
  Schuster, Shao, Shen, Small, Sohn, SooHoo, Souccar, Sun, Tazaki, Tetarenko,
  Tiede, Tilanus, Titus, Torne, Traianou, Trent, Trippe, Turk, {van Bemmel},
  {van Langevelde}, {van Rossum}, Vos, Wagner, {Ward-Thompson}, Wardle,
  Weintroub, Wex, Wharton, Wielgus, Wiik, Witzel, Wondrak, Wong, Wu, Yamaguchi,
  Yoon, Young, Young, Younsi, Yuan, Yuan, Zensus, Zhang, Zhao, Zhao, Chan, Qiu,
  Ressler, \& White}]{EHTC_2022_5}
---. 2022{\natexlab{b}}, The Astrophysical Journal, 930, L16,
  \dodoi{10.3847/2041-8213/ac6672}

\bibitem[{{Event Horizon Telescope Collaboration}
  {et~al.}(2022{\natexlab{c}}){Event Horizon Telescope Collaboration}, Akiyama,
  Alberdi, Alef, Algaba, Anantua, Asada, Azulay, Bach, Baczko, Ball,
  Balokovi{\'c}, Barrett, Baub{\"o}ck, Benson, Bintley, Blackburn, Blundell,
  Bouman, Bower, Boyce, Bremer, Brinkerink, Brissenden, Britzen, Broderick,
  Broguiere, Bronzwaer, Bustamante, Byun, Carlstrom, Ceccobello, Chael, Chan,
  Chatterjee, Chatterjee, Chen, Chen, Cheng, Cho, Christian, Conroy, Conway,
  Cordes, Crawford, Crew, {Cruz-Osorio}, Cui, Davelaar, De~Laurentis, Deane,
  Dempsey, Desvignes, Dexter, Dhruv, Doeleman, Dougal, Dzib, Eatough, Emami,
  Falcke, Farah, Fish, Fomalont, Ford, {Fraga-Encinas}, Freeman, Friberg,
  Fromm, Fuentes, Galison, Gammie, Garc{\'i}a, Gentaz, Georgiev, Goddi, Gold,
  {G{\'o}mez-Ruiz}, G{\'o}mez, Gu, Gurwell, Hada, Haggard, Haworth, Hecht,
  Hesper, Heumann, Ho, Ho, Honma, Huang, Huang, Hughes, Ikeda, Impellizzeri,
  Inoue, Issaoun, James, Jannuzi, Janssen, Jeter, Jiang, {Jim{\'e}nez-Rosales},
  Johnson, Jorstad, Joshi, Jung, Karami, Karuppusamy, Kawashima, Keating,
  Kettenis, Kim, Kim, Kim, Kim, Kino, Koay, Kocherlakota, Kofuji, Koch, Koyama,
  Kramer, Kramer, Krichbaum, Kuo, La~Bella, Lauer, Lee, Lee, Leung, Levis, Li,
  Lico, Lindahl, Lindqvist, Lisakov, Liu, Liu, Liuzzo, Lo, Lobanov, Loinard,
  Lonsdale, Lu, Mao, Marchili, Markoff, Marrone, Marscher, {Mart{\'i}-Vidal},
  Matsushita, Matthews, Medeiros, Menten, Michalik, Mizuno, Mizuno, Moran,
  Moriyama, Moscibrodzka, M{\"u}ller, Mus, Musoke, Myserlis, Nadolski, Nagai,
  Nagar, Nakamura, Narayan, Narayanan, Natarajan, Nathanail, Fuentes, Neilsen,
  Neri, Ni, Noutsos, Nowak, Oh, Okino, Olivares, {Ortiz-Le{\'o}n}, Oyama,
  {\"O}zel, Palumbo, Paraschos, Park, Parsons, Patel, Pen, Pesce, Pi{\'e}tu,
  Plambeck, PopStefanija, Porth, P{\"o}tzl, Prather, {Preciado-L{\'o}pez},
  Psaltis, Pu, Ramakrishnan, Rao, Rawlings, Raymond, Rezzolla, Ricarte,
  Ripperda, Roelofs, Rogers, Ros, {Romero-Ca{\~n}izales}, Roshanineshat,
  Rottmann, Roy, Ruiz, Ruszczyk, Rygl, S{\'a}nchez,
  {S{\'a}nchez-Arg{\"u}elles}, {S{\'a}nchez-Portal}, Sasada, Satapathy,
  Savolainen, Schloerb, Schonfeld, Schuster, Shao, Shen, Small, Sohn, SooHoo,
  Souccar, Sun, Tazaki, Tetarenko, Tiede, Tilanus, Titus, Torne, Traianou,
  Trent, Trippe, Turk, {van Bemmel}, {van Langevelde}, {van Rossum}, Vos,
  Wagner, {Ward-Thompson}, Wardle, Weintroub, Wex, Wharton, Wielgus, Wiik,
  Witzel, Wondrak, Wong, Wu, Yamaguchi, Yoon, Young, Young, Younsi, Yuan, Yuan,
  Zensus, Zhang, Zhao, \& Zhao}]{EHTC_2022_6}
---. 2022{\natexlab{c}}, The Astrophysical Journal, 930, L17,
  \dodoi{10.3847/2041-8213/ac6756}

\bibitem[{Gammie {et~al.}(2003)Gammie, McKinney, \&
  T{\'o}th}]{gammie_2003_harm}
Gammie, C.~F., McKinney, J.~C., \& T{\'o}th, G. 2003, ApJ, 589, 444,
  \dodoi{10.1086/374594}

\bibitem[{Ghavamian {et~al.}(2007)Ghavamian, Laming, \&
  Rakowski}]{ghavamian_2007_tpte_clshocks}
Ghavamian, P., Laming, J.~M., \& Rakowski, C.~E. 2007, The Astrophysical
  Journal, 654, L69, \dodoi{10.1086/510740}

\bibitem[{Goldreich \& Sridhar(1995)}]{goldreich_1995_alfvturb}
Goldreich, P., \& Sridhar, S. 1995, The Astrophysical Journal, 438, 763,
  \dodoi{10.1086/175121}

\bibitem[{Hawley(2000)}]{hawley_2000_GlobalMagnetohydrodynamicalSimulations}
Hawley, J.~F. 2000, The Astrophysical Journal, 528, 462, \dodoi{10.1086/308180}

\bibitem[{Himwich {et~al.}(2020)Himwich, Johnson, Lupsasca, \&
  Strominger}]{himwich_2020_UniversalPolarimetricSignatures}
Himwich, E., Johnson, M.~D., Lupsasca, A., \& Strominger, A. 2020, Physical
  Review D, 101, 084020, \dodoi{10.1103/PhysRevD.101.084020}

\bibitem[{Howes(2010)}]{howes_2010_PrescriptionTurbulentHeating}
Howes, G.~G. 2010, Monthly Notices of the Royal Astronomical Society, 409,
  L104, \dodoi{10.1111/j.1745-3933.2010.00958.x}

\bibitem[{Ichimaru(1977)}]{ichimaru_1977_BimodalBehaviorAccretion}
Ichimaru, S. 1977, The Astrophysical Journal, 214, 840, \dodoi{10.1086/155314}

\bibitem[{Igumenshchev {et~al.}(2003)Igumenshchev, Narayan, \&
  Abramowicz}]{igumenshchev_2003_ThreedimensionalMagnetohydrodynamicSimulations}
Igumenshchev, I.~V., Narayan, R., \& Abramowicz, M.~A. 2003, The Astrophysical
  Journal, 592, 1042, \dodoi{10.1086/375769}

\bibitem[{{Isenberg}(2004)}]{Isenberg2004}
{Isenberg}, P.~A. 2004, \jgr, 109, A03101

\bibitem[{{Isenberg} \& {Vasquez}(2007)}]{IsenbergVasquez2007}
{Isenberg}, P.~A., \& {Vasquez}, B.~J. 2007, \apj, 668, 546

\bibitem[{{Isenberg} \& {Vasquez}(2011)}]{IsenbergVasquez2011}
---. 2011, \apj, 731, 88

\bibitem[{Kawazura {et~al.}(2020)Kawazura, Schekochihin, Barnes, TenBarge,
  Tong, Klein, \& Dorland}]{kawazura_2020_IonElectronHeating}
Kawazura, Y., Schekochihin, A.~A., Barnes, M., {et~al.} 2020, Physical Review
  X, 10, 041050, \dodoi{10.1103/PhysRevX.10.041050}

\bibitem[{{Kennel} \& {Engelmann}(1966)}]{KennelEngelmann1966}
{Kennel}, C.~F., \& {Engelmann}, F. 1966, \pfluid, 9, 2377

\bibitem[{{Klein} {et~al.}(2020){Klein}, {Howes}, {TenBarge}, \&
  {Valentini}}]{Klein2020}
{Klein}, K.~G., {Howes}, G.~G., {TenBarge}, J.~M., \& {Valentini}, F. 2020,
  \jpp, 86, 905860402

\bibitem[{Kolmogorov(1941)}]{kolmogorov_1941_turbulence}
Kolmogorov, A. 1941, Doklady Akademiia Nauk SSSR, 30, 301

\bibitem[{Kunz {et~al.}(2014)Kunz, Schekochihin, \&
  Stone}]{kunz_2014_FirehoseMirrorInstabilities}
Kunz, M.~W., Schekochihin, A.~A., \& Stone, J.~M. 2014, Physical Review
  Letters, 112, 205003, \dodoi{10.1103/PhysRevLett.112.205003}

\bibitem[{{Ley} {et~al.}(2022){Ley}, {Zweibel}, {Riquelme}, {Sironi}, {Miller},
  \& {Tran}}]{Ley2022}
{Ley}, F., {Zweibel}, E.~G., {Riquelme}, M., {et~al.} 2022, arXiv:2209.00019

\bibitem[{Mahadevan \&
  Quataert(1997)}]{mahadevan_1997_AreParticlesAdvectiondominated}
Mahadevan, R., \& Quataert, E. 1997, ApJ, 490, 605, \dodoi{10.1086/304908}

\bibitem[{Marszewski {et~al.}(2021)Marszewski, Prather, Joshi, Pandya, \&
  Gammie}]{marszewski_2021_UpdatedTransferCoefficients}
Marszewski, A., Prather, B.~S., Joshi, A.~V., Pandya, A., \& Gammie, C.~F.
  2021, The Astrophysical Journal, 921, 17, \dodoi{10.3847/1538-4357/ac1b28}

\bibitem[{{McChesney} {et~al.}(1987){McChesney}, {Stern}, \&
  {Bellan}}]{McChesney1987}
{McChesney}, J.~M., {Stern}, R.~A., \& {Bellan}, P.~M. 1987, \prl, 59, 1436

\bibitem[{McKinney \& Gammie(2004)}]{mckinney_2004_harmbz}
McKinney, J.~C., \& Gammie, C.~F. 2004, The Astrophysical Journal, 611, 977,
  \dodoi{10.1086/422244}

\bibitem[{{Melville} {et~al.}(2016){Melville}, {Schekochihin}, \&
  {Kunz}}]{Melville2016}
{Melville}, S., {Schekochihin}, A.~A., \& {Kunz}, M.~W. 2016, \mnras, 459, 2701

\bibitem[{{Meyrand} {et~al.}(2021){Meyrand}, {Squire}, {Schekochihin}, \&
  {Dorland}}]{Meyrand2021}
{Meyrand}, R., {Squire}, J., {Schekochihin}, A.~A., \& {Dorland}, W. 2021,
  \jpp, 87, 535870301

\bibitem[{Mondal \& Basu(2020)}]{mondal_2020_shocksaccretion}
Mondal, S., \& Basu, P. 2020, Monthly Notices of the Royal Astronomical
  Society, 497, 2119, \dodoi{10.1093/mnras/staa2035}

\bibitem[{Mo{\'s}cibrodzka {et~al.}(2016)Mo{\'s}cibrodzka, Falcke, \&
  Shiokawa}]{moscibrodzka_2016_rhigh}
Mo{\'s}cibrodzka, M., Falcke, H., \& Shiokawa, H. 2016, Astronomy and
  Astrophysics, 586, A38, \dodoi{10.1051/0004-6361/201526630}

\bibitem[{Mo{\'s}cibrodzka \& Gammie(2018)}]{moscibrodzka_2018_ipole}
Mo{\'s}cibrodzka, M., \& Gammie, C.~F. 2018, Monthly Notices of the Royal
  Astronomical Society, 475, 43, \dodoi{10.1093/mnras/stx3162}

\bibitem[{Narayan {et~al.}(2003)Narayan, Igumenshchev, \&
  Abramowicz}]{narayan_2003_MagneticallyArrestedDisk}
Narayan, R., Igumenshchev, I.~V., \& Abramowicz, M.~A. 2003, Publications of
  the Astronomical Society of Japan, 55, L69, \dodoi{10.1093/pasj/55.6.L69}

\bibitem[{Narayan {et~al.}(2012)Narayan, S{\"a}dowski, Penna, \&
  Kulkarni}]{narayan_2012_sane}
Narayan, R., S{\"a}dowski, A., Penna, R.~F., \& Kulkarni, A.~K. 2012, Monthly
  Notices of the Royal Astronomical Society, 426, 3241,
  \dodoi{10.1111/j.1365-2966.2012.22002.x}

\bibitem[{Narayan \& Yi(1995)}]{narayan_1995_ADAF}
Narayan, R., \& Yi, I. 1995, The Astrophysical Journal, 452, 710,
  \dodoi{10.1086/176343}

\bibitem[{{\"O}zel {et~al.}(2000){\"O}zel, Psaltis, \&
  Narayan}]{ozel_2000_HybridThermalNonthermalSynchrotron}
{\"O}zel, F., Psaltis, D., \& Narayan, R. 2000, The Astrophysical Journal,
  Volume 541, Issue 1, pp. 234-249., 541, 234, \dodoi{10.1086/309396}

\bibitem[{Palumbo \& Wong(2022)}]{palumbo_2022_photonringbeta2}
Palumbo, D. C.~M., \& Wong, G.~N. 2022, The Astrophysical Journal, 929, 49,
  \dodoi{10.3847/1538-4357/ac59b4}

\bibitem[{Palumbo {et~al.}(2020)Palumbo, Wong, \&
  Prather}]{palumbo_2020_DiscriminatingAccretionStates}
Palumbo, D. C.~M., Wong, G.~N., \& Prather, B.~S. 2020, The Astrophysical
  Journal, 894, 156, \dodoi{10.3847/1538-4357/ab86ac}

\bibitem[{Prather {et~al.}(2021)Prather, Wong, Dhruv, Ryan, Dolence, Ressler,
  \& Gammie}]{prather_2021_iharm}
Prather, B., Wong, G., Dhruv, V., {et~al.} 2021, The Journal of Open Source
  Software, 6, 3336, \dodoi{10.21105/joss.03336}

\bibitem[{Psaltis {et~al.}(2020)Psaltis, Medeiros, Christian, {\"O}zel,
  Akiyama, Alberdi, Alef, Asada, Azulay, Ball, Balokovi{\'c}, Barrett, Bintley,
  Blackburn, Boland, Bower, Bremer, Brinkerink, Brissenden, Britzen, Broguiere,
  Bronzwaer, Byun, Carlstrom, Chael, Chan, Chatterjee, Chatterjee, Chen, Chen,
  Cho, Conway, Cordes, Crew, Cui, Davelaar, De~Laurentis, Deane, Dempsey,
  Desvignes, Dexter, Eatough, Falcke, Fish, Fomalont, {Fraga-Encinas}, Friberg,
  Fromm, Gammie, Garc{\'i}a, Gentaz, Goddi, G{\'o}mez, Gu, Gurwell, Hada,
  Hesper, Ho, Ho, Honma, Huang, Huang, Hughes, Inoue, Issaoun, James, Jannuzi,
  Janssen, Jiang, {Jimenez-Rosales}, Johnson, Jorstad, Jung, Karami,
  Karuppusamy, Kawashima, Keating, Kettenis, Kim, Kim, Kim, Kino, Koay, Koch,
  Koyama, Kramer, Kramer, Krichbaum, Kuo, Lauer, Lee, Li, Li, Lindqvist, Lico,
  Liu, Liu, Liuzzo, Lo, Lobanov, Lonsdale, Lu, Mao, Markoff, Marrone, Marscher,
  {Mart{\'i}-Vidal}, Matsushita, Mizuno, Mizuno, Moran, Moriyama, Moscibrodzka,
  M{\"u}ller, Musoke, Mus~Mej{\'i}as, Nagai, Nagar, Narayan, Narayanan,
  Natarajan, Neri, Noutsos, Okino, Olivares, Oyama, Palumbo, Park, Patel, Pen,
  Pi{\'e}tu, Plambeck, PopStefanija, Prather, {Preciado-L{\'o}pez},
  Ramakrishnan, Rao, Rawlings, Raymond, Ripperda, Roelofs, Rogers, Ros, Rose,
  Roshanineshat, Rottmann, Roy, Ruszczyk, Ryan, Rygl, S{\'a}nchez,
  {S{\'a}nchez-Arguelles}, Sasada, Savolainen, Schloerb, Schuster, Shao, Shen,
  Small, Sohn, SooHoo, Tazaki, Tilanus, Titus, Torne, Trent, Traianou, Trippe,
  {van Bemmel}, {van Langevelde}, {van Rossum}, Wagner, Wardle,
  {Ward-Thompson}, Weintroub, Wex, Wharton, Wielgus, Wong, Wu, Yoon, Young,
  Young, Younsi, Yuan, Yuan, Zhao, \& {EHT
  Collaboration}}]{psaltis_2020_gravtesteht}
Psaltis, D., Medeiros, L., Christian, P., {et~al.} 2020, Physical Review
  Letters, 125, 141104, \dodoi{10.1103/PhysRevLett.125.141104}

\bibitem[{Quataert(1998)}]{quataert_1998_ParticleHeatingAlfvenic}
Quataert, E. 1998, The Astrophysical Journal, 500, 978, \dodoi{10.1086/305770}

\bibitem[{Rees {et~al.}(1982)Rees, Begelman, Blandford, \&
  Phinney}]{rees_1982_IonsupportedToriOrigin}
Rees, M.~J., Begelman, M.~C., Blandford, R.~D., \& Phinney, E.~S. 1982, Nature,
  295, 17, \dodoi{10.1038/295017a0}

\bibitem[{Ressler {et~al.}(2015)Ressler, Tchekhovskoy, Quataert, Chandra, \&
  Gammie}]{ressler_2015_ElectronThermodynamicsGRMHD}
Ressler, S.~M., Tchekhovskoy, A., Quataert, E., Chandra, M., \& Gammie, C.~F.
  2015, Monthly Notices of the Royal Astronomical Society, 454, 1848,
  \dodoi{10.1093/mnras/stv2084}

\bibitem[{Reynolds {et~al.}(1996)Reynolds, Di~Matteo, Fabian, Hwang, \&
  Canizares}]{reynolds_1996_QuiescentBlackHole}
Reynolds, C.~S., Di~Matteo, T., Fabian, A.~C., Hwang, U., \& Canizares, C.~R.
  1996, Monthly Notices of the Royal Astronomical Society, 283, L111,
  \dodoi{10.1093/mnras/283.4.L111}

\bibitem[{{Riquelme} {et~al.}(2016){Riquelme}, {Quataert}, \&
  {Verscharen}}]{Riquelme2016}
{Riquelme}, M.~A., {Quataert}, E., \& {Verscharen}, D. 2016, \apj, 824, 123

\bibitem[{Rowan {et~al.}(2017)Rowan, Sironi, \&
  Narayan}]{rowan_2017_ElectronProtonHeating}
Rowan, M.~E., Sironi, L., \& Narayan, R. 2017, The Astrophysical Journal, 850,
  29, \dodoi{10.3847/1538-4357/aa9380}

\bibitem[{Rowan {et~al.}(2019)Rowan, Sironi, \&
  Narayan}]{rowan_2019_ElectronProtonHeating}
---. 2019, The Astrophysical Journal, 873, 2, \dodoi{10.3847/1538-4357/ab03d7}

\bibitem[{Ryan {et~al.}(2018)Ryan, Ressler, Dolence, Gammie, \&
  Quataert}]{ryan_2018_TwotemperatureGRRMHDSimulations}
Ryan, B.~R., Ressler, S.~M., Dolence, J.~C., Gammie, C., \& Quataert, E. 2018,
  The Astrophysical Journal, 864, 126, \dodoi{10.3847/1538-4357/aad73a}

\bibitem[{S{\k{a}}dowski {et~al.}(2013)S{\k{a}}dowski, Narayan, Penna, \&
  Zhu}]{sadowski_2013_EnergyMomentumMass}
S{\k{a}}dowski, A., Narayan, R., Penna, R., \& Zhu, Y. 2013, Monthly Notices of
  the Royal Astronomical Society, 436, 3856, \dodoi{10.1093/mnras/stt1881}

\bibitem[{S{\k{a}}dowski {et~al.}(2017)S{\k{a}}dowski, Wielgus, Narayan,
  Abarca, McKinney, \&
  Chael}]{sadowski_2017_RadiativeTwotemperatureSimulations}
S{\k{a}}dowski, A., Wielgus, M., Narayan, R., {et~al.} 2017, Monthly Notices of
  the Royal Astronomical Society, 466, 705, \dodoi{10.1093/mnras/stw3116}

\bibitem[{{Schekochihin}(2022)}]{schekochihin_2022_review}
{Schekochihin}, A.~A. 2022, Journal of Plasma Physics, 88, 155880501,
  \dodoi{10.1017/S0022377822000721}

\bibitem[{Shapiro {et~al.}(1976)Shapiro, Lightman, \&
  Eardley}]{shapiro_1976_twotemp}
Shapiro, S.~L., Lightman, A.~P., \& Eardley, D.~M. 1976, The Astrophysical
  Journal, 204, 187, \dodoi{10.1086/154162}

\bibitem[{Sironi \& Beloborodov(2020)}]{sironi_2020_KineticReconnection}
Sironi, L., \& Beloborodov, A.~M. 2020, The Astrophysical Journal, 899, 52,
  \dodoi{10.3847/1538-4357/aba622}

\bibitem[{Sironi \& Narayan(2015)}]{sironi_2015_ElectronHeatingIon}
Sironi, L., \& Narayan, R. 2015, The Astrophysical Journal, 800, 88,
  \dodoi{10.1088/0004-637X/800/2/88}

\bibitem[{{Squire} {et~al.}(2023){Squire}, {Kunz}, {Arzamasskiy}, {Johnston},
  {Quataert}, \& {Schekochihin}}]{Squire2023}
{Squire}, J., {Kunz}, M.~W., {Arzamasskiy}, L., {et~al.} 2023, arXiv:2303.00468

\bibitem[{{Squire} {et~al.}(2017){Squire}, {Kunz}, {Quataert}, \&
  {Schekochihin}}]{Squire2017}
{Squire}, J., {Kunz}, M.~W., {Quataert}, E., \& {Schekochihin}, A.~A. 2017,
  \prl, 119, 155101

\bibitem[{{Squire} {et~al.}(2022){Squire}, {Meyrand}, {Kunz}, {Arzamasskiy},
  {Schekochihin}, \& {Quataert}}]{Squire2022}
{Squire}, J., {Meyrand}, R., {Kunz}, M.~W., {et~al.} 2022, \astrnat

\bibitem[{Tidman \& Krall(1971)}]{tidman_1971_collisionlessshocks}
Tidman, D.~A., \& Krall, N.~A. 1971, Shock Waves in Collisionless Plasmas

\bibitem[{Tran \& Sironi(2020)}]{tran_2020_eheating_lowbetashocks}
Tran, A., \& Sironi, L. 2020, The Astrophysical Journal, 900, L36,
  \dodoi{10.3847/2041-8213/abb19c}

\bibitem[{{Tran} {et~al.}(2022){Tran}, {Sironi}, {Ley}, {Zweibel}, \&
  {Riquelme}}]{Tran2022}
{Tran}, A., {Sironi}, L., {Ley}, F., {Zweibel}, E.~G., \& {Riquelme}, M.~A.
  2022, arXiv:2209.12902

\bibitem[{Walker {et~al.}(2018)Walker, Hardee, Davies, Ly, \&
  Junor}]{walker_2018_m87jet}
Walker, R.~C., Hardee, P.~E., Davies, F.~B., Ly, C., \& Junor, W. 2018, The
  Astrophysical Journal, 855, 128, \dodoi{10.3847/1538-4357/aaafcc}

\bibitem[{Werner \& Uzdensky(2021)}]{werner_2021_3dReconnectionPairs}
Werner, G.~R., \& Uzdensky, D.~A. 2021, Journal of Plasma Physics, 87,
  905870613, \dodoi{10.1017/S0022377821001185}

\bibitem[{Werner {et~al.}(2018)Werner, Uzdensky, Begelman, Cerutti, \&
  Nalewajko}]{werner_2018_NonthermalParticleAcceleration}
Werner, G.~R., Uzdensky, D.~A., Begelman, M.~C., Cerutti, B., \& Nalewajko, K.
  2018, Monthly Notices of the Royal Astronomical Society, 473, 4840,
  \dodoi{10.1093/mnras/stx2530}

\bibitem[{Wong \& Gammie(2022)}]{wong_2022_helium}
Wong, G.~N., \& Gammie, C.~F. 2022, The Astrophysical Journal, 937, 60,
  \dodoi{10.3847/1538-4357/ac854d}

\bibitem[{Wong {et~al.}(2022)Wong, Prather, Dhruv, Ryan, Mo{\'s}cibrodzka,
  Chan, Joshi, Yarza, Ricarte, Shiokawa, Dolence, Noble, McKinney, \&
  Gammie}]{wong_2022_PATOKA}
Wong, G.~N., Prather, B.~S., Dhruv, V., {et~al.} 2022, The Astrophysical
  Journal Supplement Series, 259, 64, \dodoi{10.3847/1538-4365/ac582e}

\bibitem[{Yuan \& Narayan(2014)}]{yuan_2014_HotAccretionFlows}
Yuan, F., \& Narayan, R. 2014, Annual Review of Astronomy and Astrophysics, 52,
  529, \dodoi{10.1146/annurev-astro-082812-141003}

\bibitem[{Zhdankin {et~al.}(2021)Zhdankin, Uzdensky, \&
  Kunz}]{zhdankin_2021_ProductionPersistenceExtreme}
Zhdankin, V., Uzdensky, D.~A., \& Kunz, M.~W. 2021, The Astrophysical Journal,
  908, 71, \dodoi{10.3847/1538-4357/abcf31}

\bibitem[{Zhdankin {et~al.}(2018)Zhdankin, Uzdensky, Werner, \&
  Begelman}]{zhdankin_2018_NumericalInvestigationKinetic}
Zhdankin, V., Uzdensky, D.~A., Werner, G.~R., \& Begelman, M.~C. 2018, Monthly
  Notices of the Royal Astronomical Society, 474, 2514,
  \dodoi{10.1093/mnras/stx2883}

\bibitem[{Zhdankin {et~al.}(2019)Zhdankin, Uzdensky, Werner, \&
  Begelman}]{zhdankin_2019_ElectronIonEnergization}
---. 2019, Physical Review Letters, 122, 055101,
  \dodoi{10.1103/PhysRevLett.122.055101}

\end{thebibliography}
\bibliographystyle{aasjournal}

\end{document}